\documentclass[11pt]{article}

\usepackage{a4wide}

\usepackage{amsmath,amssymb,amsfonts} 

\usepackage{graphics,graphicx}
\usepackage[mathscr]{eucal}

\def\bb{\begin{eqnarray}}
\def\ee{\end{eqnarray}}

\newtheorem{definition}{Definition}[section]
\newtheorem{lemma}{Lemma}[section]

\newtheorem{proposition}{Proposition}[section]

\newcommand{\Proof}{\begin{proof}}
\newcommand{\QED}{\end{proof} \noindent}

\begin{document}
${}$
\begin{center}

{\Large Quantum physics, fields and closed timelike curves: \\[6pt]
The D-CTC condition in quantum field theory}
\\[20pt]
{\large {\bf J\"urgen Tolksdorf}${}^{(a)}$ and {\bf Rainer Verch}${}^{(b)}$}
\\[16pt]
${}^{(a)}$ Max-Planck-Institute for Mathematics in the Sciences, 04109 Leipzig, Germany
\\[10pt]
${}^{(b)}$ Institute for Theoretical Physics, University of Leipzig, 04009 Leipzig, Germany

\end{center}
${}$ \\ \\ 
${}$ \hfill {\sl Dedicated to the memory of Rudolf Haag}
\\ \\
{\bf Abstract} The D-CTC condition is a condition originally proposed by David Deutsch as 
a condition on states of a quantum communication network that contains ``backward time-steps''
in some of its branches. It has been argued that this is an analogue for quantum processes
in the presence of closed timelike curves (CTCs). The unusual properties of states of
quantum communication networks that fulfill the D-CTC condition have been discussed extensively
in recent literature. In this work, the D-CTC condition is investigated in the framework 
of quantum field theory in the local, operator-algebraic approach due to Haag and Kastler.
It is shown that the D-CTC condition cannot be fulfilled in states which are analytic for 
the energy, or satisfy the Reeh-Schlieder property, for a certain class of processes and
initial conditions.
On the other hand, if a quantum field theory admits sufficiently many uncorrelated states
across acausally related spacetime regions (as implied by the split property), then the 
D-CTC condition can always be fulfilled approximately to arbitrary precision. As this result
pertains to quantum field theory on globally hyperbolic spacetimes where CTCs are absent,
one may conclude that interpreting the D-CTC condition as characteristic for quantum processes
in the presence of CTCs could be misleading, and should be regarded with caution.
Furthermore, a construction of the quantized
massless Klein-Gordon field on the Politzer spacetime, often viewed as spacetime analogue for
quantum communication networks with backward time-steps, is proposed in this work.


\section{Introduction}
\setcounter{equation}{0}
The theoretical possiblities of physics in the presence of closed timelike curves have recently acquired considerable
attention in the context of the physical foundations of quantum computing (see, e.g., 
\cite{AhnMyersRalphMann,BrunWW,PienaarRalphMyers,RingbauerEtAl}). 
The theoretical basis
of these investigations goes back to a seminal article by David Deutsch \cite{Deutsch} suggesting that
the effect of closed timelike curves (CTCs) on the dynamics of 
quantum systems can be simulated by quantum networks with built-in ``backward time-steps'' in some branches of the networks.
Thereby, issues of spacetime structure and CTCs in the sense of general relativity are deliberately put aside.
It is then argued in \cite{Deutsch} that, at least as far as networks with simple computing processes are concerned,
if the states carried along the network are classical (``bits''), then the presence of ``backward time-steps'' (analoguos
to CTCs) imposes strong consistency conditions on the possible states of the networks, to the effect that only
``trivial'' computing processes are possible, while in the case where the states carried along the networks are 
quantum mechanical (``qubits''), non-trivial computing processes are always possible in the presence
of ``backward time-steps'', in the sense that there are always
quantum states that are compatible with the ensuing consistency constraints. 
In recent publications, experiments in support of that proposition have also been presented \cite{LloydetAlEx,RingbauerEtAl}. The authors 
of \cite{RingbauerEtAl} even
posit that ``quantum mechanics therefore allows for causality violation without paradoxes whilst remaining consistent with
relativity''. (See also \cite{Deutsch} for a related statement, and \cite{Dunlap} for some recent critizism.
There is by now some amount of literature on the approach initiated in \cite{Deutsch} which
we shall not be reviewing here; for further discussion, see the articles cited before as well as 
\cite{BubStairs,LloydEtAl,LloydetAlEx,EarmanSmeenkWuethrich} and references
cited there.)

Investigations on quantum networks with ``backward time-steps'' do not refer to
the spacetime structure of general relativity and therefore the relation between
the quantum network setting and quantum physics in the
presence of CTCs within the framework of general relativity is not entirely clear --- see however some
discussion in \cite{Deutsch} on this point as well as the articles \cite{Politzer,GoldwPPThorne,FewsHiguWells} and also 
further below in this introduction ---  consequently,
it is not completely settled if the quantum processes in  networks involving ``backward time-steps'' are merely an
analogy to (but in fact, in significant operational terms, different from) 
quantum processes in the presence of CTCs in the spacetime sense. A discussion of certain aspects of that 
issue is the topic of the present work. 

The basis of our discussion is the operator algebraic approach to quantum field theory, or more generally ``local quantum physics'',
as initially laid out by Haag and Kastler \cite{HaagKastler} and further developed by Haag and several collaborators -- see
the monograph \cite{Haag} for an outline and relevant references. The central point in this approach is to endow the 
observables of a quantum system with a locality concept that relates to where and when in spacetime the measurements corresponding
to the observables are performed. Furthermore, the causality structure of spacetime localization of observables is 
translated into algebraic relations. This is an operational approach, in contrast to attempts of understanding
quantum physics in the presence of CTCs in terms of many-world interpretations of quantum physics as in \cite{Deutsch}
(see also \cite{Dunlap} and references cited there).

In the model-independent setting of operator-algebraic quantum field theory 
according to Haag and Kastler (\cite{Haag}, see also \cite{Wald-QFTCST,RecAdvQFT} and literature cited there for 
generalizations to curved spacetimes)
a quantum field theory on a (globally hyperbolic) spacetime manifold $M$ with Lorentzian
metric $g \equiv g_{ab}$ is described by a family of $C^*$ algebras $\{\mathcal{A}(O)\}_{O \subset M}$
indexed by open spacetime regions $O$ that have compact closure in $M$.\footnote{In certain
cases, it may be appropriate to (i) drop the requirement that $O$ have
compact closure in $M$, or (ii) to impose further requirements on the spacetime regions
$O$ in order to be considered as localization regions for observables,
such as simple connectedness or geometric regularity conditions. To keep the
discussion simple at this point, we will not consider such details here.} 
The idea is that
$\mathcal{A}(O)$ contains (in mathematical idealization) the observables of the system which can be measured 
within the spacetime region $O$. As an expression of that, there are two essential conditions imposed
on $\{\mathcal{A}(O)\}_{O \subset M}$: {\it Isotony}, demanding that $\mathcal{A}(O_1) \subset 
\mathcal{A}(O_2)$ if $O_1 \subset O_2$, and {\it Locality}, requiring that the algebras
$\mathcal{A}(O_1)$ and $\mathcal{A}(O_2)$ commute elementwise if the spacetime regions
$O_1$ and $O_2$ are acausally related, meaning that they cannot be connected by a smooth
causal curve in $M$. 
 Supposing that there is a unique $C^\ast$
algebra $\mathcal{A}$ generated by all the $\mathcal{A}(O)$, a state of the system
is a linear functional $\omega: \mathcal{A} \to \mathbb{C}$ subject to the conditions
of positivity ($\omega({\bf a}^*{\bf a}) \ge 0$ for all ${\bf a} \in \mathcal{A}$)
and normalization  ($\omega({\bf 1}) = 1$ where ${\bf 1}$ is the unit element in
$\mathcal{A}$, assumed to coincide with the unit element of any $\mathcal{A}(O)$).
In more suggestive fashion, one might write
$\langle {\bf a} \rangle_\omega = \omega({\bf a})$ for ${\bf a} \in \mathcal{A}$
which emphasizes that a state is nothing but
an expectation value functional. Typically, the set of all states mathematically defined by positivity
and normalization is too large and one needs criteria which select states which, intuitively,
correspond to physical situations with finite particle- and energy-densities. On Minkowski spacetime,
this is achieved by considering states that can be obtained by (localized) operations from
a vacuum state; for quantum field theories in curved spacetime, the microlocal spectrum
condition plays an analogous role in serving as selection criterion for physical states \cite{Radzikowski,
Wald-QFTCST,Fewster-Verch-Review,KhavMor,HollandsWald-Review}. 

The operator algebraic approach has been very succesful in the mathematical and conceptual analysis
of quantum field theories, also for quantum field theories in curved spacetimes --- however, mostly
under the assumption that the underlying spacetimes are {\it globally hyperbolic} and that the
quantum field theory is {\it arbitrarily localizable}. A globally hyperbolic spacetime contains
Cauchy-surfaces and this implies absence of CTCs in such spacetimes (with this property being
stable under small variations of the spacetime metric) \cite{Wald-book,Bernal-Sanchez}. A quantum
field theory given as $\{\mathcal{A}(O)\}_{O \subset M}$ is arbitrarily localizable if the 
$\mathcal{A}(O)$ are ``large'' (infinite dimensional) even if $O$ becomes arbitrarily small, i.e.\
concentrated around a spacetime point. A more specific way stating it is to say that {\it additivity} holds
for $\{\mathcal{A}(O)\}_{O \subset M}$, meaning that any $\mathcal{A}(O)$ is contained in the algebra generated by the 
$\mathcal{A}(O_j)$ whenever the $\{O_j\}$ form a cover of $O$. Theories of this type typically are 
{\it locally covariant}, so that the structure of the local algebras is independent of any ambient
globally hyperbolic spacetime. In the recent years, the concept of local covariant quantum field 
theories has led to significant advances in quantum field theory in curved spacetimes \cite{BFV,Fewster-Verch-Review,
HollandsWald-Review,RecAdvQFT}.

Within the operator algebraic approach, there has also been some progress in the understanding
of conformal quantum field theories in the presence of spacetime boundaries \cite{Rehren-Longo} and
quantum field theories on spacetimes with CTCs such as AdS \cite{Rehren,Bu-Florig-Sum}. It has also been noticed
earlier that some spacetimes with CTCs can be regarded as having a periodically ``rolled-up'' time-axis,
and that constistent QFTs can be constructed for such spacetimes by ``un-rolling'' the time-variable and
imposing appropriate periodicity conditions \cite{AvisIshamStory,FewsHiguchi}.
However, it seems that for more general spacetimes $M$
with CTCs, it is unclear if quantum field theories can be constructed in terms of 
$\{\mathcal{A}(O)\}_{O \subset M}$ which comply with the conditions of isotony and locality
and are arbitrarily localizable (so in particular, the local algebras $\mathcal{A}(O)$ are ``large'') and
admit a large set of states fulfilling suitable regularity conditions such as the microlocal spectrum
condition known from quantum field theory on globally hyperbolic spacetimes. In fact, it has been
conjectured that this cannot be done under certain assumptions on
the spacetime structure \cite{Hawking}\footnote{An important assumption
made in \cite{Hawking} and \cite{KayRadWald} is that there is a 
{\it compactly generated Cauchy horizon} separating the globally hyperbolic part
of spacetime from the part containing CTCs}, and some results corroborating that conjecture
have been derived in \cite{KayRadWald} and \cite{KayFlocality}. However, these results rest 
on very specific assumptions on quantum field theories on spacetimes containing CTCs, especially their
arbitrary localizability, and it might well turn out that this is not an appropriate generalization
of quantum field theory to spacetimes with CTCs. 

The question as to what principles should be invoked
to characterize physics on spacetimes containing CTCs is intricate \cite{Hawking,KayFlocality,EarmanSmeenkWuethrich}.
It might well be that an answer to that question will involve radical departures from our current
picture of physical processes in spacetime, as has been suggested some time ago by Rudolf Haag related to 
the physical content of quantum gravity \cite{Haag1990a,Haag1990b,Haag1995}. In such a situation, the approach by
David Deutsch \cite{Deutsch} could provide a new vantage point on the said question. In fact, in 
\cite{Deutsch} and some other works \cite{Politzer,GoldwPPThorne,FewsHiguWells}, relations between the quantum network approach
and quantum (field) theories on a certain type of spacetimes with CTCs have been investigated. We will now summarize
some pieces of that discussion. 

As argued in \cite{Deutsch}, the key element of a quantum network simulating processes in the presence
of CTCs is of the following form, as depicted in Fig.\ 1:
\begin{center}
 \includegraphics[width=10.5cm]{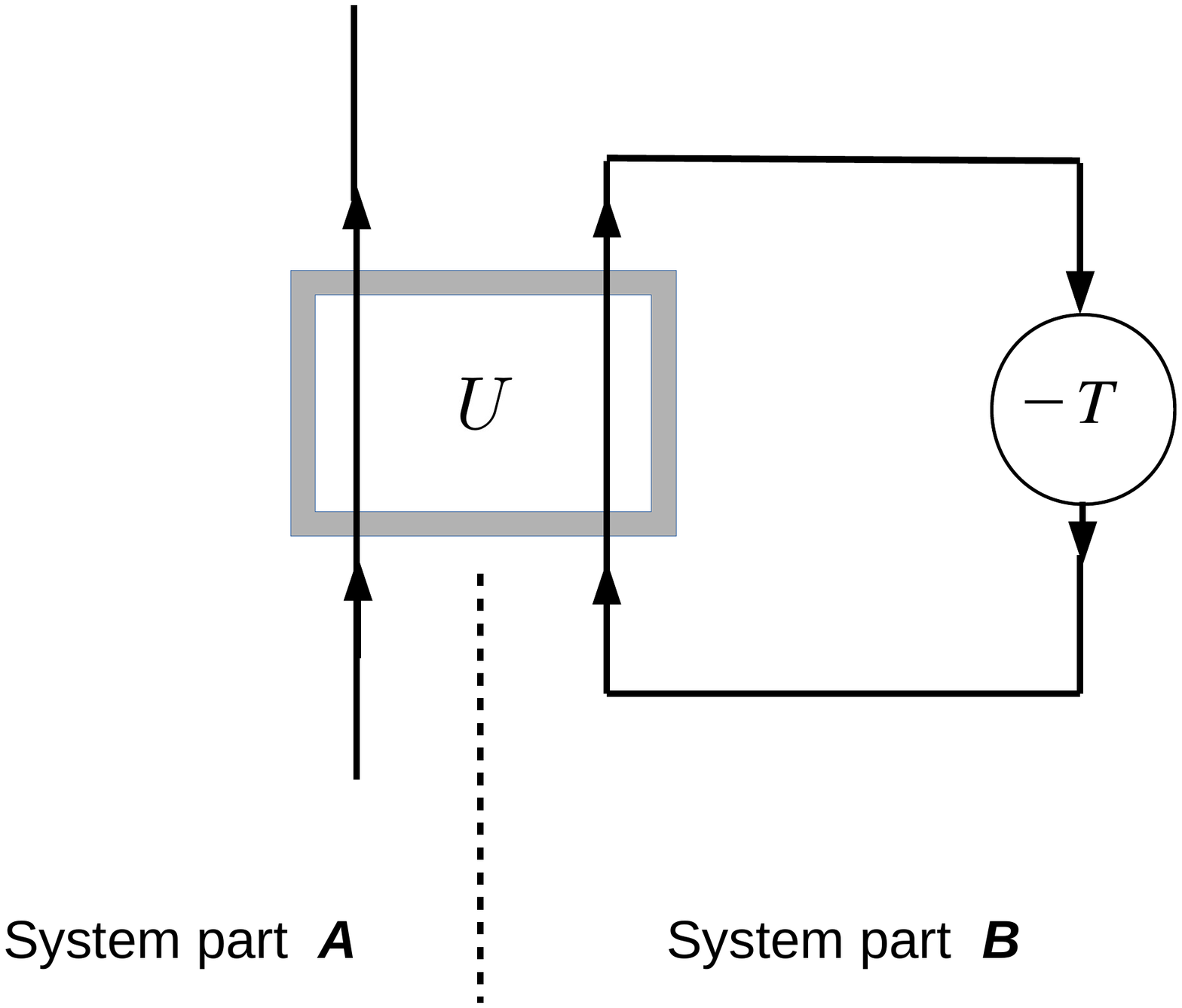}
\end{center}
{\small {\bf Figure 1}. Network of two quantum systems where the right hand side part is subject to a 
``backward time-step'', symbolized by $-T$ (taken from \cite{Deutsch}). See discussion in the text.}
\\[8pt]
In this quantum network, one has two branches, denoted by solid lines with arrows, where the direction
of the arrows indicates the evolution of quantum systems (e.g., ``qubits'') along the branches in time. 
The left branch, corresponding to ``System part $A$'', evolves normally in time, without any CTC-related
effects. It interacts with the right branch, corresponding to ``System part $B$'', through a unitary
operator $U$ which can be viewed as a dynamical evolution operator, or scattering operator: The 
quantum systems enter at some ``early time'' into the interaction, symbolized by the $U$-box, from below,
and leave the interaction, in a $U$-transformed state, at some ``later time'', at the 
top end of the $U$-box. The duration of the interaction is assumed to be $T$ in suitable units. However,
on the right hand side branch, the quantum system in the state after the $U$-interaction is fed into the
interaction again, after undergoing a ``backward time-step'', symbolized by the encircled $-T$. In that manner,
the dynamics of the quantum system along the right hand side branch is taken to follow a CTC \cite{Deutsch}.

The effect of $-T$ in the right hand side branch of the network is described by viewing
Fig.\ 1  as illustrating the dynamical evolution of a bi-partite quantum system.
Following \cite{Deutsch}, 
Hilbert spaces of quantum states $\mathcal{H}_A$ and $\mathcal{H}_B$, taken to be finite dimensional, are
associated with the $A$ and $B$ system parts. 
The observable algebras are, respectively, ${\sf B}(\mathcal{H}_A)$ and ${\sf B}(\mathcal{H}_B)$ where
we use the notation ${\sf B}(\mathcal{H})$ for the set of bounded linear operators on the Hilbert space
$\mathcal{H}$. The full bi-partite system then has the Hilbert space of quantum states $\tilde{\mathcal{H}} = \mathcal{H}_A
\otimes \mathcal{H}_B$ with observable algebra ${\sf B}(\tilde{\mathcal{H}}) = 
{\sf B}(\mathcal{H}_A) \otimes {\sf B}(\mathcal{H}_B)$. The dynamical evolution coupling the two system 
parts is described by a unitary $U \in {\sf B}(\tilde{\mathcal{H}})$.
(At this stage, $U$ is completely generic.)
The unitary $U$ takes an initially given
state $\varrho_{\rm ini} = \varrho_A \otimes \varrho_B$, where $\varrho_A$ and $\varrho_B$ are density matrices
on $\mathcal{H}_A$ and $\mathcal{H}_B$, respectively, to a final state $\varrho_{\rm fin} =
U^\ast \varrho_A \otimes \varrho_B U$. In \cite{Deutsch}, Deutsch has suggested to describe  
 the ``backward time-step'' $-T$ in 
the $B$-part branch by the condition that, if the system is ``initially'' in a state 
$\tilde{\varrho} = \varrho_A \otimes \varrho_B$, then 
\begin{equation} \label{dc-trversion}
  \varrho_B = {\rm Tr}_A(U^\ast \varrho_A \otimes \varrho_B U)
\end{equation}
must be fulfilled,
where ${\rm Tr}_A$ denotes the partial trace over the $\mathcal{H}_A$ part of $\mathcal{H}_A \otimes
\mathcal{H}_B$.  This means that the $\varrho_B$-state in which
the $B$-part of the system had been initially prepared returns to itself after the coupled dynamics with the $A$-part of
the system has taken effect.  It has been shown in \cite{Deutsch} that for any given unitary 
$U$ on $\mathcal{H}_A \otimes \mathcal{H}_B$ and any density matrix $\varrho_A$ on $\mathcal{H}_A$,
there is a density matrix $\varrho_B$ on $\mathcal{H}_B$ such that \eqref{dc-trversion} holds (in general,
the solution will be non-unique). Deutsch and several other authors have taken this result as indicating
that it may well be possible that quantum processes are consistent in the presence of
CTCs while analogous classical processes are inconsistent. This is illustrated in \cite{Deutsch} by the example of
a quantum circuit with a CNOT gate (playing the role of $U$) coupling two qubit systems,
as opposed to classical bit-systems. Whence, condition \eqref{dc-trversion} has come to be
referred to as {\it Deutsch condition} for quantum processes in the presence of CTCs, or {\it D-CTC condition}, for short.

In \cite{Deutsch} and \cite{Politzer} (see also \cite{GoldwPPThorne,FewsHiguWells}), a situation analogous
to the network of Fig.\ 1 has been considered in terms of a quantum system propagating on a particular
two-dimensional spacetime containing CTCs, the {\it Politzer spacetime} introduced in \cite{Politzer}. 
This spacetime is, somewhat informally, described by Fig.\ 2: 
\begin{center}
\includegraphics[width=11cm]{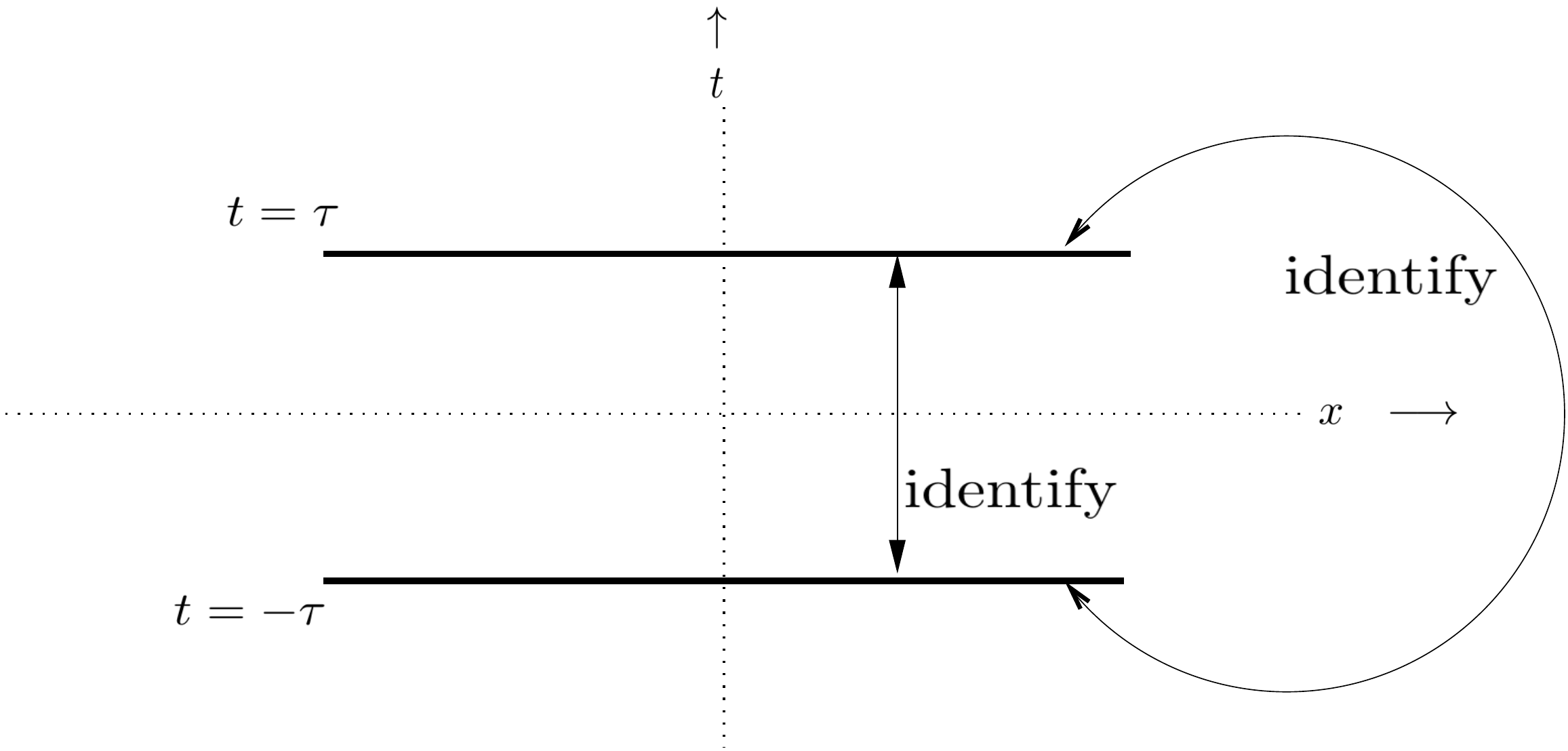}
\end{center}
{\small {\bf Figure 2}. Sketch of the Politzer spacetime. The line segments $S_\pm$ described in the text are represented by
the solid horizontal lines.}
\\[8pt]
In 2-dimensional Minowski spacetime with time-coordinate $t$ and space-coordinate $x$, the two spacelike
line segments $S_{\pm} = \{ (t,x) : t = \pm \tau, \ |x| \le L\}$ (for some chosen $\tau,L > 0$)
are cut away, and then one identifies each point $(\tau - \varepsilon,x)$ with the point $(-\tau + \varepsilon,x)$
in the limit $\varepsilon \to 0+$ ($|x| \le L$), and one also identifies each point 
$(\tau + \varepsilon,x)$ with the point $(-\tau - \varepsilon,x)$ in the limit $\varepsilon \to 0+$ $(|x| \le L)$. 
As already noted in \cite{Politzer}, this construction does not readily render a two-dimensional
smooth manifold because one would have to specify in some detail the manifold structure at the 
points $(\pm \tau,\pm L)$. Apart from that difficulty, away from these critical points the resulting set of spacetime points 
can be given a manifold structure
and can be equipped with the flat Minkowski coordinate metric. One may then consider ``wave functions''
$\psi(t,x)$ on the Politzer spacetime which may take values in some $\mathbb{C}^N$ and are subject
to an equation of motion of the form
\begin{equation} \label{theDeqn}
 \partial_t \psi(t,x) = D_x \psi(t,x)
\end{equation}
where $D_x$ denotes a diffential operator acting with respect to the $x$-coordinates. Upon making appropriate
choices of $N$ and $D_x$, this covers the cases of the Schr\"odinger equation, the wave equation (transformed into
a first order system)
or the Dirac equation. 

On 2-dim.\ Minkowski spacetime, one has well-posedness of the initial value problem for these
cases, and there is a dynamical evolution operator $U$ which carries the initial values $\psi_{-\tau}(x) = \psi(-\tau,x)$ 
of any solution $\psi(t,x)$ to \eqref{theDeqn} on the 
equal time hypersurface $\Sigma_- = \{ (-\tau,x) : x \in \mathbb{R} \}$ to the initial values $\psi_{\tau}(x) = \psi(\tau,x)$
on the equal time hypersurface $\Sigma_+ = \{ (\tau,x) : x \in \mathbb{R} \}$. The initial data spaces can be equipped with
a suitable Hilbert space structure so that $U$ becomes unitary. 

Similarly one can try to construct such a dynamical evolution operator $U$ carrying initial data of solutions
to \eqref{theDeqn} on $\Sigma_-$ to the data on $\Sigma_+$ (interpreting the hypersurfaces, respectively, as 
$\{ \lim_{\varepsilon \to 0+}\, (\pm \tau \mp \varepsilon,x) : |x| \le L\}$) for the Politzer spacetime. 
In this case, there is the additional constraint 
\begin{equation} \label{initdataconstraint}
 \lim_{\varepsilon \to 0+}\, \psi(\tau - \varepsilon,x) = \lim_{\varepsilon \to 0+}\, \psi(-\tau + \varepsilon,x) \quad
 (|x| \le L)
\end{equation}
that solutions to \eqref{theDeqn} need to fulfill. Some results on the existence of such a dynamical evolution operator $U$
for solutions $\psi(t,x)$ to \eqref{theDeqn} on (various versions of) the Politzer spacetime are contained in the references
\cite{Politzer,GoldwPPThorne,FewsHiguWells}. A relation to the network setting depicted by Fig.\ 1 can be obtained as follows:
 ``System part $A$'' is identified with solutions $\psi(t,x)$ to \eqref{theDeqn} having initial data on $\Sigma_-$ that 
have support in $\{ \lim_{\varepsilon \to 0+}\, (-\tau + \varepsilon,x) : |x| > L\}$ while ``System part $B$'' corresponds to
solutions with initial data on $\Sigma_-$ having support in $\{ \lim_{\varepsilon \to 0+}\, (-\tau + \varepsilon,x) : |x| \le L\}$.
The interaction, or dynamical coupling $U$ in Fig.\ 1 corresponds to the dynamical evolution of initial data from $\Sigma_-$ to $\Sigma_+$;
this introduces a dynamical coupling between the system parts in the sense that the supports of 
initial data of solutions to \eqref{theDeqn} will tend to spread 
in as time evolves. The Deutsch condition \eqref{dc-trversion} then has the initial data constraint \eqref{initdataconstraint} as its counterpart.

We will address some further aspects of quantum fields on the Politzer spacetime in more detail in Sec.\ 4. The main concern
of this article is to put the Deutsch condition \eqref{dc-trversion} into the context of operator algebraic quantum field theory on
{\it globally hyperbolic} spacetime and to see under which conditions it can be fulfilled, or not fulfilled. That is to say, we
will start from an arbitrarily localizable quantum field theory $\{\mathcal{A}(O)\}_{O \subset M}$ on some globally hyperbolic 
spacetime $M$, where we assume that the local algebras are actually von Neumann algebras in some Hilbert space representation, so that
$\mathcal{A}(O) \subset {\sf B}(\mathcal{H})$ for some Hilbert space $\mathcal{H}$. Then we will specify ``System part $A$'' and 
``System part $B$'' by choosing as their respective observable algebras the local algebras $\mathcal{A}(O_A)$ and $\mathcal{A}(O_B)$
where $O_A$ and $O_B$ are two acausally related spacetime regions in $M$, as shown in Fig.\ 3.
\begin{center}
 \includegraphics[width=9.5cm]{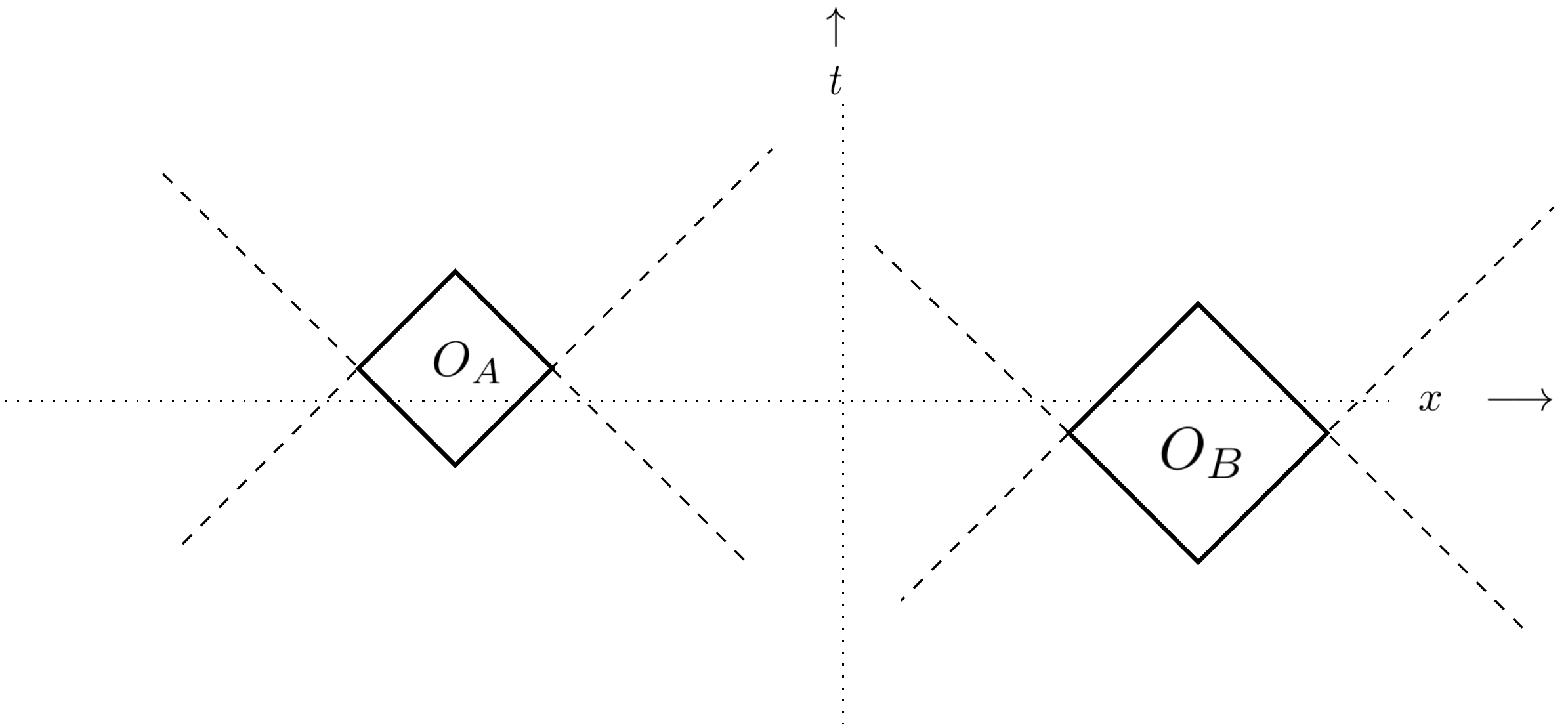}
\end{center}
{\small {\bf Figure 3}. Two acausally related regions $O_A$ and $O_B$. Their regions of causal influence are bounded by light rays,
represented by dashed lines at $45^\circ$ inclination.}
\\[8pt]
The problem of fulfilling the 
Deutsch condition \eqref{dc-trversion} can be posed as follows in the present general setting: Given a 
unitary $U \in {\sf B}(\mathcal{H})$ and a normal state (a density
matrix state) $\omega_A$ on $\mathcal{A}(O_A)$, is there a normal state $\tilde{\omega}$ on ${\sf B}(\mathcal{H})$ such that 
its partial state on $\mathcal{A}(O_A)$ coincides with $\omega_A$ and such that the partial states of $\tilde{\omega}$ and
$\tilde{\omega}(U^\ast\,.\,U)$ on $\mathcal{A}(O_B)$ coincide, i.e.\ such that
\begin{equation} \label{DC}
 \tilde{\omega}({\bf a}) = \omega_A({\bf a})\,, \quad \text{and} \quad  \tilde{\omega}({\bf b}) = \tilde{\omega}(U^\ast{\bf b}U)
\end{equation}
holds for all ${\bf a} \in \mathcal{A}(O_A)$ and all ${\bf b} \in \mathcal{A}(O_B)$\,?

We will establish complementary answers to this question, based on generic assumptions in quantum field theory. At this
point, we describe our results informally; full details will be given in the following chapters.
\\[6pt]
(I) It will first be asumed that the spacetime is stationary, i.e.\ possesses a time-symmetry. Then, if $\tilde{\omega}$
fulfills \eqref{DC} for any given $\omega_A$ such that $\tilde{\omega}$ and $\tilde{\omega}(U^\ast\,.\,U)$ fulfill
an {\it analyticity condition} with respect to the time-symmetry, we show that $\tilde{\omega}$ and 
$\tilde{\omega}(U^\ast \,.\, U)$ must coincide on the full algebra of observables $\mathcal{A}$.
Furthermore, assuming that $\tilde{\omega}$ fulfills a {\it Reeh-Schlieder property}, and a certain
bound on the difference of correlations between the algebras $\mathcal{A}(O_A)$ and $\mathcal{A}(O_B)$
in the states $\tilde{\omega}$ and $\tilde{\omega}(U^\ast \,.\, U)$, we show that $U$ must commute with
all operators in $\mathcal{A}(O_B)$. This holds without having to assume the spacetime to be stationary.
 Very roughly, these results state that, if $U$ induces a non-trivial coupling
between $\mathcal{A}(O_A)$ and $\mathcal{A}(O_B)$, then there can be no state $\tilde{\omega}$ fulfilling
\eqref{DC} which shows a certain strong form of acausal entanglement.
\\[6pt]
(II) Here we consider quantum field theory on a generic globally hyperbolic spacetime.
It will be assumed that any pair of normal states $\omega_A$ on
on $\mathcal{A}(O_A)$ and $\omega_B$ on $\mathcal{A}(O_B)$ can be extended to a normal state on the
full algebra $\mathcal{A}$ (which is always possible in quantum field theories obeying the {\it
split property}, generically assumed to be fulfilled for physically realistic quantum field models).
Then given any unitary operator $U$ on $\mathcal{H}$
and any normal state $\omega_A$ on $\mathcal{A}(O_A)$, there are normal states $\tilde{\omega}$ that fulfill \eqref{DC}
approximately, i.e.\ given any $\varepsilon > 0$ and $R > 0$, there is a normal state $\tilde{\omega}$ such that
\begin{equation} \label{DCepsilon}
\tilde{\omega}({\bf a}) = \omega_A({\bf a})\,, \quad \text{and}
\quad  |\tilde{\omega}({\bf b}) - \tilde{\omega}(U^\ast{\bf b}U)| < \varepsilon
\end{equation}
holds for all ${\bf a} \in \mathcal{A}(O_A)$ and ${\bf b} \in \mathcal{A}(O_B)$ with $|| {\bf b} || \le R$.
This result shows that under general assumptions in quantum field theory, the D-CTC condition \eqref{DC} can
always be fulfilled to arbitrary precision.
\\[6pt]
In Section 2, we will describe the operator algebraic setting of quantum field theory in globally
hyperbolic and stationary spacetimes in more detail, as far as needed for our purposes. 
We then demonstrate our results in Section 3. 
In Section 4, we propose a construction of the quantized real, massless Klein-Gordon field on the
Politzer spacetime. To our knowledge, this is the first construction of a quantum field theory
on the Politzer spacetime.

There will be a Conclusion in Section 5. Section 6 is an Appendix.

\section{Quantum field theory on Lorentzian spacetimes}

\setcounter{equation}{0}

\subsection{Spacetime structure}

To begin, we very briefly summarize some notions related to spacetimes in the sense of general relativity.
By a {\it spacetime}, we mean a $d$-dimensional ($d \ge 2$) smooth, connected manifold $M$ together with a 
Lorentzian metric $g$; we choose the metric signature as $(+\ - \,\ldots\, -)$. Then we will often write
$\boldsymbol{M} = (M,g)$ to denote a spacetime. We will also assume that spacetimes are time-orientable,
i.e.\ that there is a smooth, timelike vector field on $M$. A time-orientation and an orientation are assumed to be chosen but
we will usually not display it in the notation for a spacetime unless needed. 

In the discussion of causality in spacetime, it is useful to introduce, for any subset $S$ of a given spacetime $(M,g)$,
the {\it causal future$(+)$/past$(-)$ sets} $J^\pm(S)$. They are defined, respectively, as the subsets of $M$
with can be reached by a future-directed $(+)$/past-directed $(-)$ causal curve emanating from $S$. Of interest
is also the {\it domain of dependence} $D(S)$ of $S$: A point $q$ in $M$ is contained in $D(S)$ if 
all past-directed or all future-directed causal curves which start at $q$ and cannot be
past/future extended in $M$ intersect $S$. (For the precise definition of future/past non-extendibility,
or endpoint-freeness, of causal curves, we refer to the literature.) Thus, roughly speaking, the causal
set $J(S) = J^+(S) \cup J^-(S)$ contains all points which can be causally influenced from any point in $S$,
and $D(S)$ contains all points which are causally determined from all points in $S$ taken together. 
Given a subset $S$ of $M$, we define its (open) {\it causal complement} in $\boldsymbol{M}$ as
\begin{equation}
  S^\perp = M \backslash \overline{J(S)} \,.
\end{equation}
One can also give the definition of a {\it closed timelike curve} in a spacetime $(M,g)$: This is 
a smooth timelike curve, defined on a real interval, such that the curve runs through at least one
point in spacetime more than once, i.e.\ arrives at the same point in spacetime at different curve
parameter values. In more formal notation, a smooth curve $\gamma: I \to M$, where $I$ is a (non-trivial)
real interval, is a CTC if it is timelike, i.e.\ the tangent $\dot{\gamma}(t)$ is a timelike
vector for all $t \in I$, and if there are parameters $t_1 < t_2$ in $I$ such that $\gamma(t_1) = 
\gamma(t_2)$. 

Of interest for physics are, in particular, globally hyperbolic spacetimes. A spacetime $(M,g)$ is
called {\it globally hyperbolic} if it contains Cauchy-surfaces. A Cauchy-surface is a smooth
hypersurface $S$ in $M$ such that each non-extendable causal curve in $(M,g)$ intersects $S$
exactly once. This can also be expressed as $D(S) = M$. If a spacetime is globally hyperbolic, then
there are many Cauchy-surfaces; in fact, it is then possible to foliate the spacetime smoothly 
into Cauchy-surfaces. That means, there are a ``model Cauchy-surface'' $\tilde{S}$ (a $d-1$-dimensional
smooth manifold) and a diffeomorphism $F : \mathbb{R} \times \tilde{S} \to M$ such that
for all $t \in \mathbb{R}$, $S_t = F(\{t\} \times \tilde{S})$ is a Cauchy-surface \cite{Wald-book,Bernal-Sanchez}.
If a spacetime $(M,g)$ is globally hyperbolic, then its causal structure is very regular; there
are no CTCs in $(M,g)$, and the absence of CTCs is also stable against small perturbations
of the spacetime metric $g$ \cite{Wald-book}. Furthermore, many examples of physically relevant spacetimes
which arise as solutions to Einstein's equations are globally hyperbolic. 

A globally hyperbolic spacetime will be called {\it stationary} if there is a foliation $F : \mathbb{R} \times \tilde{S} \to M$
so that the time-shifts $\vartheta_{t} : F(t',q) \to F(t' + t,q)$ $(t' \in \mathbb{R}\,, \ q \in \tilde{S})$ form
a smooth, 1-parametric group of isometries of the spacetime with respect to $t \in \mathbb{R}$ \cite{Wald-book}.


\subsection{Algebraic quantum field theory on a general spacetime}

Let $\boldsymbol{M} = (M,g)$ be a spacetime (not necessarily globally hyperbolic). 
In the operator algebraic approach, a quantum field theory on $\boldsymbol{M}$ consists of  
a family $\{ \mathcal{A}(O)\}_{O \in \mathcal{K}(\boldsymbol{M})}$ of $C^*$ algebras; 
$\mathcal{K}(\boldsymbol{M})$ is a set of open subsets of $M$ whose precise specification
is part of the particular quantum field theory. For the case that $\boldsymbol{M}$ is globally
hyperbolic, there are examples of quantum field theories where $\mathcal{K}(\boldsymbol{M})$
is the set of open, relatively compact subsets of $M$; this includes the quantized linear fields
such as the Klein-Gordon field, the Dirac field or the electromagnetic field, and perturbatively
constructed self-interacting fields. In these cases, the algebras $\mathcal{A}(O)$ are still ``large''
even when $O$ is ``small'', and in a sense which can be made precise (after passing to 
suitably regular Hilbert space representations), the algebras $\mathcal{A}(O)$ are generated by
bounded functions of field operators $\Phi(f)$ where the test-functions $f$ have support in $O$.
Therefore, one may call quantum field theories of this kind {\it arbitrarily localizable}. Usually, the
$\mathcal{A}(O)$ are taken as algebras of observables localized within the spacetime region $O$ and
therefore, as already mentioned in the introduction, then
the family $\{ \mathcal{A}(O) \}_{O \in \mathcal{K}(\boldsymbol{M})}$
is to fulfill the conditions of isotony, $\mathcal{A}(O_1) \subset \mathcal{A}(O_2)$ if $O_1 \subset O_2$,
and locality, $[\mathcal{A}(O_1),\mathcal{A}(O_2)] = \{ 0 \}$ if $O_1 \subset O_2^\perp$. The 
just mentioned examples of arbitrarily localizable quantum field theories fulfill these conditions (for the 
Dirac field, one needs to define the observable algebras appropriately). 

Without going into too much detail at this point, we mention that in cases where $\boldsymbol{M}$ is
not globally hyperbolic, e.g.\ if $\boldsymbol{M}$ contains CTCs, there is reason to expect that the setting
ought to be generalized. One point is that allowing $\mathcal{K}(\boldsymbol{M})$ to consist of
all open, relatively compact subsets $O$ of $M$ could result in trivial $\mathcal{A}(O)$ for possibly
a large set of $O$s: The presence of CTCs induces constraints on solutions to hyperbolic
partial differential equations on a spacetime which might constitute a hinderance to their localization,
i.e.\ their vanishing on causal complements of subsets $O$ of the spacetime. As observables in 
$\mathcal{A}(O)$ are usually
built through ``quantizing'' such localized solutions to hyperbolic equations, this may result in very sparse algebras
$\mathcal{A}(O)$ particularly for ``small'' spacetime regions $O$. Therefore, in order that 
the $\mathcal{A}(O)$ are ``large'', in the presence of CTCs one ought to expect that $\mathcal{K}(\boldsymbol{M})$
consists of ``extended'' subsets of $M$, and possibly only such which are of a particular type or ``shape''.
That is to say, one ought to envisage that quantum field theories on spacetimes with CTCs are not arbitrarily
localizable. Furthermore, it is well possible that in the presence 
of CTCs the isotony condition assumes the more complicated form $\varphi_{O_1,O_2} : \mathcal{A}(O_1) \to 
\mathcal{A}(O_2)$ for $O_1 \subset O_2$ with a cocycle $\varphi_{-,-}$ of algebraic embeddings; possibly,
even more complicated embedding relations might appear. Similarly, the locality condition might have
to be generalized, e.g.\ implemented at the level of (suitably defined) Hilbert space representations
as in \cite{GLRV}, or by changing the concept of causal complement as in \cite{Bu-Florig-Sum}.
``Infinitely extended'' localization regions for observables appear also in Rehren's approach to
the AdS-CFT correspondence \cite{Rehren} and in quantum field theories with infinite spin
\cite{LongoMoriRehren}. 
Having mentioned the possiblity that assuming quantum field theories on spacetimes with CTCs 
to be arbitrarily localizable may turn out to be too restrictive and to result in theories
having very few observables, we will now limit ourselves to considering quantum field theories
on globally hyperbolic spacetimes under the assumption of their being arbitrarily localizable.
In fact, the progress achieved in quantum field theory in curved spacetimes over the recent years
based on the concept of {\it local covariant quantum field theory} \cite{BFV,Fewster-Verch-Review,HollandsWald-Review} uses
arbitrarily localizable quantum field theories as starting point. 

\subsection{Arbitrarily localizable quantum field theory on globally hyperbolic spacetimes}

For an arbitrarily localizable quantum field theory $\{ \mathcal{A}(O) \}_{\mathcal{K}(\boldsymbol{M})}$
on a globally hyperbolic spacetime $\boldsymbol{M}$ it is typically no restriction to assume that 
all the $\mathcal{A}(O)$ are $C^\ast$-subalgebras of a global $C^\ast$-algebra $\mathcal{A}(\boldsymbol{M})$
which is generated by all those subalgebras. We will henceforth adopt this assumption, mainly
for convenience. Then we recall \cite{BratRobins} that any state $\omega: \mathcal{A}(\boldsymbol{M}) \to \mathbb{C}$
induces a Hilbert space representation  $(\mathcal{H}_\omega,\pi_\omega,\Omega_\omega)$, called {\it GNS representation},
where $\mathcal{H}_\omega$ is a Hilbert space, $\Omega_\omega$ is a unit vector in $\mathcal{H}_\omega$, and
$\pi_\omega$
is a unital $\ast$-representation of $\mathcal{A}(\boldsymbol{M})$ by bounded linear operators on $\mathcal{H}_\omega$,
such that the following two properties hold: (i) $\Omega_\omega$  
is cyclic for $\pi_\omega$\,, i.e.\ $\pi_\omega(\mathcal{A})\Omega_\omega$
is dense in $\mathcal{H}_\omega$, and (ii) 
$\omega({\bf a}) = \langle \Omega_\omega, \pi_\omega({\bf a}) \Omega_\omega \rangle$  $({\bf a} \in
  \mathcal{A})$\,,
where $\langle \chi,\psi \rangle$ is the scalar product of vectors $\chi,\psi \in \mathcal{H}_\omega$.
 
We have already pointed out that selection criteria must be supplied to pick out physical states
for a quantum field theory on a generic spacetime, and we have already mentioned that 
states fulfilling the {\it microlocal spectrum condition} are considered, by many results,
as very good candidates for physical states 
(see \cite{Radzikowski,Wald-QFTCST,Sanders-qfHad,KhavMor,Fewster-Verch-Review,HollandsWald-Review}
and references cited therein). 
For quantized linear fields,
it has been shown that any quasifree state $\omega_1$ fulfilling the microlocal spectrum condition
arises locally as a density matrix state in the GNS representation of any other such state $\omega_2$
\cite{Verch-QEq,DAnHol}. 
In other words, for quantized linear fields and quasifree states $\omega_1$ and $\omega_2$ fulfilling the
microlocal spectrum condition, there is for any open, relatively compact subset $O$ of $M$ a density matrix
$\varrho = \varrho_{O,\omega_1}$ on $\mathcal{H}_{\omega_2}$ such that
\begin{equation} \label{localnormal}
 \omega_1({\bf a}) = {\rm Tr}(\varrho \pi_{\omega_2}({\bf a})) \quad \ \ ({\bf a} \in \mathcal{A}(O))\,.
\end{equation}
This relation between states $\omega_1$ and $\omega_2$ is referred to by saying that $\omega_1$ {\it is
locally normal with respect to} $\omega_2$. 

More generally, no matter what the selection criterion is in detail, we assume that it leads to
a set of states $\mathscr{S}(\boldsymbol{M})$ on $\mathcal{A}(\boldsymbol{M})$ which forms an
equivalence class under the relation of local normality. (For convenience, we will also impose the
condition that the GNS Hilbert spaces $\mathcal{H}_\omega$ of all $\omega \in \mathscr{S}(\boldsymbol{M})$
are separable --- this is fulfilled for quantized linear fields and states fulfilling the microlocal
spectrum condition.) That is to say, for any pair of states $\omega_1,\omega_2 \in \mathscr{S}(\boldsymbol{M})$,
$\omega_1$ is locally normal with respect to $\omega_2$. In this case, when investigating properties
of states in $\mathscr{S}(\boldsymbol{M})$ locally, i.e.\ in restriction to the algebras $\mathcal{A}(O)$,
one can simply pick any state $\omega \in \mathscr{S}(\boldsymbol{M})$ and henceforth work with the
local algebras $\pi_\omega(\mathcal{A}(O))$ in the GNS Hilbert space representation of $\omega$, since
any other state $\omega'$ in $\mathscr{S}(\boldsymbol{M})$ is, in restriction to the local algebras,
given by density matrices. Therefore, one can pass to the local von Neumann algebras
\begin{equation}
 \mathcal{N}_\omega(O) = \overline{\pi_\omega(\mathcal{A}(O))}
\end{equation}
where the overlining means weak closure (convergence in the sense of expectation values). 
Then any other state $\omega' \in \mathscr{S}(\boldsymbol{M})$ extends uniquely from a state on $\mathcal{A}(O)$ (upon restriction)
to a state on $\mathcal{N}(O)$ since it is locally induced by a density matrix, cf.\ \eqref{localnormal}.

Working at the level of local von Neumann algebras has often several technical advantages in 
algebraic quantum field theory and this is also the case for our discussion. 
A von Neumann algebra $\mathcal{N}$ is a weakly closed $*$-subalgebra of ${\sf B}(\mathcal{H})$ for
some Hilbert space $\mathcal{H}$ containing the unit operator. According to von Neumann's bicommutant
theorem \cite{BratRobins}, $\mathcal{N}$ is a von Neumann algebra if and only if
\begin{equation}
 \mathcal{N} = \mathcal{N}''
\end{equation}
where, for any subset $\mathcal{L}$ of ${\sf B}(\mathcal{H})$, $\mathcal{L}'' = (\mathcal{L}')'$
is the {\it double commutant}, while the {\it commutant} is defined by
\begin{equation}
 \mathcal{L}' = \{ {\bf b} \in {\sf B}(\mathcal{H}_\omega) : {\bf ab} = {\bf ba} \ \text{for all} \ {\bf a} \in
  \mathcal{L} \}\,.
\end{equation}

\subsection{Analyticity in the energy, Reeh-Schlieder and timelike tube property} \label{AnRSttt}

Let it be assumed that $\{\mathcal{A}(O)\}_{O \in \mathcal{K}(\boldsymbol{M})}$ is an arbitrarily
localizable quantum field theory on a globally hyperbolic spacetime $\boldsymbol{M}$, that
$\mathscr{S}(\boldsymbol{M})$ is a state space which forms an equivalence class with respect
to local normality. It is also assumed that $\{\mathcal{A}(O)\}_{O \in \mathcal{K}(\boldsymbol{M})}$
fulfills additivity. 
Furthermore, we assume that the spacetime $\boldsymbol{M}$ is stationary and that $\{ \vartheta_t\}_{t \in
\mathbb{R}}$ is an associated 1-parametric group of time-shift isometries of $\boldsymbol{M}$. 
Then it is natural to assume (and actually holds for local covariant quantum field theories, in 
particular, for quantum fields obey a linear hyperbolic field equation, such as the Klein-Gordon
field or the Dirac field) that there is a 1-parametric group $\{ \alpha_t\}_{t \in \mathbb{R}}$ of
isomorphisms of $\mathcal{A}(\boldsymbol{M})$ so that 
\begin{equation} \label{timeshiftcovariance}
        \alpha_t(\mathcal{A}(O)) = \mathcal{A}(\vartheta_t O) \quad (O \in \mathcal{K}(\boldsymbol{M}),\ t \in \mathbb{R})\,.
\end{equation} 
As another assumption, we suppose that for any $\omega \in \mathscr{S}(\boldsymbol{M})$, the functions
$ t \mapsto \omega({\bf a}_1\alpha_t({\bf b}){\bf a}_2)$ are continuous for all ${\bf a}_1,{\bf a}_2,{\bf b} \in 
\mathcal{A}(\boldsymbol{M})$. Moreover, we assume that there is a {\it vacuum state} $\omega_0 \in \mathscr{S}(\boldsymbol{M})$,
characterized as having the properties: (i) $\omega_0(\alpha_t({\bf a})) = \omega_0({\bf a})$ for all $t \in \mathbb{R}$,
${\bf a} \in \mathcal{A}(\boldsymbol{M})$; (ii) $ -id/dt|_{t =0}\,\omega_0({\bf a}\alpha_t({\bf b})) \ge 0$ for all
${\bf a,b} \in \mathcal{A}(\boldsymbol{M})$; (iii) $\omega_0$ is
a pure state (not a convex combination of two or more different states). Writing the GNS-representation of $\omega_0$ as
$(\mathcal{H}_0,\pi_0,\Omega_0)$, one can express that $\omega_0$ is a ground state equivalently by demanding the following
properties to hold:
(A) There is a continuous unitary group $V_t$ $(t \in \mathbb{R})$ on $\mathcal{H}_0$ so that 
$\pi_0(\alpha_t({\bf a})) = V_t \pi_0({\bf a}) V_t^\ast$ and $V_t \Omega_0 = \Omega_0$ for all $t \in \mathbb{R}$
and ${\bf a} \in \mathcal{A}(\boldsymbol{M})$; (B) $V_t = {\rm e}^{itH}$ with a selfadjoint Hamiltonian operator $H$ in
$\mathcal{H}_0$ that has non-negative spectrum; (C) $\pi_0(\mathcal{A}(\boldsymbol{M}))'' = {\sf B}(\mathcal{H}_0)$.
For quantum field theory on Minkowski spacetime, the vacuum state is a ground state for the time-shift isometries
with respect to any inertial time-coordinate. For more general stationary spacetimes, the existence of ground states
for quantum fields obeying a linear hyperbolic field equation has been established for a wide range of stationary
spacetimes \cite{KayGround,Furlani,WMJin}. 

With these assumptions, let $\omega \in \mathscr{S}(\boldsymbol{M})$ be a density matrix state in the GNS-representation
of $\omega_0$, i.e.\ $\omega({\bf a}) = {\rm Tr}(\varrho \pi_0({\bf a})$ $({\bf a} \in \mathcal{A}(\boldsymbol{M}))$.
Then $\omega$ will we called an {\it analytic} state if for any given choice of finitely many  ${\bf a}_j$, $j = 1,\ldots,N$,
in ${\sf B}(\mathcal{H}_0)$, the functions $(t_1,\ldots,t_{N+1}) \mapsto {\rm Tr}(\varrho V_{t_1}{\bf a}_1V_{t_2}{\bf a}_2 
\cdots V_{t_N}{\bf a}_NV_{-t_{N+1}})$ are boundary values of an analytic function in $\mathbb{R}^{N+1} \times (0,\lambda)^{N+1}$
for some $\lambda > 0$. Particular examples are density matrices of the form $\varrho = Z^{-1}E \tilde{\varrho} E$ where 
$E$ is a spectral projector of $H$ corresponding to a finite energy interval,  $\tilde{\varrho}$ is any density matrix on
$\mathcal{H}_0$, and $Z = {\rm Tr}(E\tilde{\varrho})$. Density matrix states of this form are called {\it states of finite energy}. 
Correspondigly we define
a unit vector $\psi$ to be of {\it finite energy} if it lies in the range of a finite-energy interval spectral projector $E$ of
$H$, and to be analytic if $\psi$ is a strongly analytic vector for $H$ (which is always the case if $\psi$ is of finite energy).

A unit vector $\psi$ in $\mathcal{H}_0$ is said to possess the {\it Reeh-Schlieder
property} if for any $O \in \mathcal{K}(\boldsymbol{M})$, the vector $\psi$ {\it is cyclic for}
$\mathcal{N}(O)$, which means by definition,
\begin{equation}
   \mathcal{N}(O)\psi = \{ {\bf a}\psi : {\bf a} \in \mathcal{N}(O)\} \ \ \ 
   \text{is dense in} \ \, \mathcal{H}_0\,.
\end{equation}
Here, we have denoted the local von Neumann algebras in the GNS representation of the ground
state $\omega_0$ simply as $\mathcal{N}(O) = \pi_0(\mathcal{A}(O))''$, without any subscript.

Note that this is equivalent to requiring that $\psi$ {\it is separating} for the
commutant $\mathcal{N}(O)'$ of $\mathcal{N}(O)$, where a vector $\psi$ is separating
for a set of bounded operators $\mathcal{L}$ if, for any ${\bf b} \in \mathcal{L}$,
${\bf b}\psi = 0$ implies ${\bf b} = 0$. Owing to the assumption of locality on 
$\{\mathcal{A}(O)\}_{O \in \mathcal{K}(\boldsymbol{M})}$ it follows that
$\mathcal{N}(O_\times)  \subset \mathcal{N}_\omega(O)'$ whenever $O_\times \in O^ \perp$,
and hence
a vector $\psi \in \mathcal{H}_0$  that fulfills the Reeh-Schlieder property is also
separating for all $\mathcal{N}_\omega(O_\times)$ if $O_\times$ has an open causal complement. 

If the underlying spacetime $\boldsymbol{M}$ is Minkowski spacetime and if the ground state GNS-representation
considered
satisfies additionally the usual Wightman-Haag-Kaster assumption of a vacuum representation, whereby 
$\Omega_0$ is the vacuum vector and hence invariant under a continuous unitary representation of the 
group of translations on Minkowski spacetime which fulfills the relativistic spectrum condition (so that
the ground state $\omega_0$ is a common ground state with respect to the time-shifts of any time-like inertial
coordinate), then the {\it Reeh-Schlieder theorem} asserts that any unit vector $\psi$ in $\mathcal{H}_0$
which is analytic (for $H$) possesses the Reeh-Schlieder property \cite{ReehSchlieder,StreaterWightman,Haag}. 
(Our standing assumption
of additivity of $\{\mathcal{A}(O)\}_{O \in \mathcal{K}(\boldsymbol{M})}$ is of relevance for that result.)
This theorem has several
important mathematical and conceptual consequences in quantum field theory and we refer to \cite{Haag} for
discussion. The Reeh-Schlieder property leads to entanglement
across von Neumann algebras $\mathcal{N}(O_A)$ and $\mathcal{N}(O_B)$ for acausally separated
spacetime regions $O_A$ and $O_B$; see \cite{CliftonHalvorson,VerchWerner} and literature cited there.

Under the same standard assumptions for a quantum field theory on Minkowski spacetime in vacuum representation, Borchers has 
proved the {\it timelike tube theorem} which states that the von Neumann algebra generated by a ``timelike
tube region'' agrees with ${\sf B}(\mathcal{H}_0)$ \cite{Borchers-ttt}; that is,
\begin{equation}
       \left ( \bigcup_{t \in \mathbb{R}} \mathcal{N}(O + te_0) \right)'' = {\sf B}(\mathcal{H}_0) 
\end{equation}
for any $O \in \mathcal{K}(\boldsymbol{M})$ and any time-like vector $e_0$ on Minkowski spacetime.

A timelike tube theorem has also been proved in the GNS-representations of ground states for the 
quantized Klein-Gordon and Dirac fields on globally hyperbolic, stationary spacetimes \cite{Strohmaier-ttt}, in the
form 
\begin{equation} \label{ttt-prop}
       \left ( \bigcup_{t \in \mathbb{R}} \mathcal{N}(\vartheta_t O) \right)'' = {\sf B}(\mathcal{H}_0) 
\end{equation}
for any $O \in \mathcal{K}(\boldsymbol{M})$.

We put on record the following result.

\begin{lemma} \label{ttt-equality}
 Let $\{\mathcal{A}(O)\}_{O \in \mathcal{K}(\boldsymbol{M})}$ be an arbitrarily localizable quantum field theory fulfilling
 additivity on a globally hyperbolic, stationary spacetime $\boldsymbol{M}$ with 1-parametric isometry group of time-shifts
 $\{ \vartheta_t \}_{t \in \mathbb{R}}$. Assume that the timelike tube property \eqref{ttt-prop} holds in the GNS-representation
 of a chosen ground state $\omega_0$.
 
 Suppose that $\omega_1$ and $\omega_2$ are two
 states which are normal with respect $\omega_0$ (so they arise as density matrix states in the GNS representation of $\omega_0$)
 and that $\omega_1$ and $\omega_2$ are both analytic in the sense described above. Then
\begin{equation}
 \omega_1({\bf a}) = \omega_2({\bf a}) \quad ({\bf a} \in \mathcal{A}(O))
\end{equation}
 for any fixed $O \in \mathcal{K}(\boldsymbol{M})$ implies $\omega_1 = \omega_2$ on $\mathcal{A}(\boldsymbol{M})$. 
\end{lemma}
{\it Proof}. The method of proof is standard and completely follows the pattern of the proof of the Reeh-Schlieder theorem,
so it will suffice to provide just a sketch. 

Choose $N \in \mathbb{N}$. Then let $\check{O} \in \mathcal{K}(\boldsymbol{M})$ be such that $ \bigcup_{-\lambda < t' < \lambda}\vartheta_{t'}(\check{O}) \subset O$ 
with some sufficiently small $\lambda > 0$. If ${\bf a}_j \in \mathcal{A}(\check{O})$, then 
$\alpha_{t'_1}({\bf a}_1) \cdots \alpha_{t'_N}({\bf a}_N) \in \mathcal{A}(O)$ and hence,
\begin{equation}
\omega_1( \alpha_{t'_1}({\bf a}_1) \cdots \alpha_{t'_N}({\bf a}_N)) = \omega_2(\alpha_{t'_1}({\bf a}_1) \cdots \alpha_{t'_N}({\bf a}_N))
\end{equation}
for all $t'_j \in (-\lambda,\lambda)$. If $\varrho_1$ and $\varrho_2$ are the density matrices inducing the states $\omega_1$ and $\omega_2$
in the GNS representation $(\mathcal{H}_0,\pi_0,\Omega_0)$ of $\omega_0$, then the previous equation implies (writing
${\bf a}_j$ for $\pi_0({\bf a}_j)$ to simplify notation)
\begin{equation}
 {\rm Tr}((\varrho_1 - \varrho_2)( V_{t_1} {\bf a}_1 V_{t_2} {\bf a_2} \cdots V_{t_N} {\bf a}_N V_{-t_{N+1}})) = 0
\end{equation}
where $t_1 = t'_1$, $t_{N+1} = t_N'$, $t_k = t'_k - t'_{k-1}$ ($k = 2,\ldots,N$). Owing to the assumed analyticity property, the 
function of the $(t_1,\ldots,t_{N+1})$ on the left hand side is the boundary value of an analytic function, and it vanishes on 
an open set of the $(t_1,\ldots,t_{N+1})$, so it vanishes everywhere according to the edge-of-the wedge-theorem \cite{StreaterWightman}. 
Then using additivity, one can see that ${\rm Tr}(\varrho_1 {\bf b}) = {\rm Tr}(\varrho_2 {\bf b})$ for all ${\bf b} \in
\bigcup_{t \in \mathbb{R}} \mathcal{N}(\vartheta_t \check{O})$ and by the assumed timelike tube property, that 
implies ${\rm Tr}(\varrho_1 {\bf b}) = {\rm Tr}(\varrho_2 {\bf b})$
for all ${\bf b} \in {\sf B}(\mathcal{H}_0)$, proving the statement. \hfill $\Box$
\\[8pt]
We mention a further fact related to the Reeh-Schlieder property, namely that under very general assumptions,
the local von Neumann algebras $\mathcal{N}(O)$ of quantum fields on Minkwski spacetime, as well as on general 
globally hyperbolic spacetimes, are of {\it properly infinite type}, more precisely, of type ${\rm III}_1$
\cite{FredenhagenIII1,WollenbergIII1,VerchIII1}. This holds under the condition that 
$O^\perp$ is non-empty, and that the geometrical shape of $O$ satisfies some very mild regularity properties.
An important consequence is that a normal state $\omega$ on $\mathcal{N}(O)$ can be represented as by means
of a {\it state vector} in the Hilbert space $\mathcal{H}$ in which $\mathcal{N}(O)$ is defined, i.e.\ there is 
some (non-unique) unit vector $\eta_\omega \in \mathcal{H}$ such that
\begin{equation}
 \omega({\bf b}) = {\rm Tr}(\varrho_\omega {\bf b}) = \langle \eta_\omega, {\bf b} \eta_\omega \rangle \quad 
 ({\bf b} \in \mathcal{N}(O))\,.
\end{equation}
The condition that $\omega$ is a normal state on $\mathcal{N}(O)$ means, by definition, that there is 
a density matrix $\varrho_\omega$ on $\mathcal{H}$ so that the equality just stated is fulfilled; but in particular,
it is possible to present the normal state $\omega$ on $\mathcal{N}(O)$ even as expectation value in the 
unit vector $\eta_\omega$ \cite{BratRobins,Haag}. 
The assumption that $O$ possesses a non-void causal complement is important here, as it ensures that $\mathcal{N}(O)'$ is
large. As a further consequence, no normal state on $\mathcal{N}(O)$ is a pure state.

\subsection{Extension of states and split property}

As in the previous subsection, it will be assumed that $\{ \mathcal{A}(O) \}_{O \in \mathcal{K}(\boldsymbol{M})}$
is an arbitrarily localizable quantum field theory on a globally hyperbolic spacetime $\boldsymbol{M}$, and
that $\mathscr{S}(\boldsymbol{M})$ is a set of states on $\mathcal{A}(\boldsymbol{M})$ 
which forms an equivalence class with respect to local normality.

Let $\omega \in \mathscr{S}(\boldsymbol{M})$, and denote by
$\{ \mathcal{N}(O)\}_{O \in \mathcal{K}(\boldsymbol{M})}$ the family 
of local von Neumann algebras in the GNS representation of $\omega$.
Suppose that $O_A$ and $O_B$ are two {\it properly} acausally related spacetime regions
in $\mathcal{K}(\boldsymbol{M})$, which means that both $\overline{O_A}$ and $\overline{O_B}$
possess open neighbourhoods that are still acausally related. In this case, one may pose the
following extension problem: Given two normal states $\omega_A$ and $\omega_B$ on $\mathcal{N}_A = \mathcal{N}(O_A)$
and $\mathcal{N}_B = \mathcal{N}(O_B)$, respectively, do these arise as partial states from a 
single normal state on ${\sf B}(\mathcal{H}_\omega)$\,?

We formalize that issue in the following definition.

\begin{definition}
 Let $\mathcal{N}_A$ and $\mathcal{N}_B$ a pair of von Neumann algebras on a Hilbert space
$\mathcal{H}$ with $\mathcal{N}_A \subset \mathcal{N}_B'$. We say that a pair of 
normal states $\omega_A$ on $\mathcal{N}_A$ and $\omega_B$ on $\mathcal{N}_B$ is 
{\em normal state extendable} if there is a normal state $\tilde{\omega}$ on
${\sf B}(\mathcal{H})$ such that 
\begin{equation}
 \tilde{\omega}({\bf a}) = \omega_A({\bf a}) \quad \text{and} \quad  \tilde{\omega}({\bf b}) = \omega_B({\bf b})
\end{equation}
holds for all ${\bf a} \in \mathcal{N}_A$ and all ${\bf b} \in \mathcal{N}_B$. We then
say that $\tilde{\omega}$ is  
a {\em normal state extending} $\omega_A$ and $\omega_B$
and we define ${\rm NEx}(\omega_A,\omega_B)$ as the set of all normal states extending
$\omega_A$ and $\omega_B$. Again, we adopt the convention to define ${\rm NEx}(\omega_A,\omega_B)$ only
for the case that $\omega_A$ and $\omega_B$ are normal state extendable 
so that ${\rm NEx}(\omega_A,\omega_B)$ is, according to convention, always non-empty.
\end{definition}

In quantum field theory, it is of considerable interest if an extending state may even be chosen
correlation free, that is, as a product state. Using the same notation as in the previous definition,
let us say that
a pair of 
normal states $\omega_A$ on $\mathcal{N}_A$ and $\omega_B$ on $\mathcal{N}_B$ is 
{\em normal product state extendable} if there is a normal state $\tilde{\omega}$ on
${\sf B}(\mathcal{H})$ such that 
\begin{equation}
 \tilde{\omega}({\bf a}{\bf b}) = \omega_A({\bf a}) \omega_B({\bf b})
\end{equation}
holds for all ${\bf a} \in \mathcal{N}_A$ and all ${\bf b} \in \mathcal{N}_B$. Correspondingly,
we define
 $\tilde{\omega}$ as a 
{\em normal product state extending} $\omega_A$ and $\omega_B$
and  ${\rm NPEx}(\omega_A,\omega_B)$ as the set of all normal product states extending
$\omega_A$ and $\omega_B$. 

Obviously, a product state cannot have the Reeh-Schlieder property. It is a very interesting
fact that, while in quantum field theory there is typically a dense set of state vectors
in the GNS representation of any state $\omega \in \mathscr{S}(\boldsymbol{M})$ that fulfill
the Reeh-Schlieder property, one can establish in many quantum field theoretical models 
in Minkowski spacetime as well as in globally hyperbolic spacetime that ${\rm NPEx}(\omega_A,\omega_B)$
is non-empty for any choice of normal states $\omega_A$ and $\omega_B$ on $\mathcal{N}_A = \mathcal{N}(O_A)$
and $\mathcal{N}_B = \mathcal{N}(O_B)$ whenever $O_A$ and $O_B$ are properly acausally related.
For further discussion we refer to \cite{BuSplit,Verch-Nuc,Haag,DAnHol,FewsterSplit} and to the reviews
\cite{Summers-Indep,Fewster-Verch-Review}. 

Moreover, we mention the connection between product state extendability and the
{\it split property} for the family local von Neumann algebras $\{ \mathcal{N}(O)\}_{O \in \mathcal{K}(\boldsymbol{M})}$
arising in the GNS representation of any $\omega \in \mathscr{S}(\boldsymbol{M})$. The split property
states that for any pair 
 $O_1,O_2 \in \mathcal{K}(\mathcal{M})$
with $\overline{D(O_1)} \subset O_2$ there is a type I factor von Neumann algebra $\mathcal{B}$ on
$\mathcal{H}_\omega$ so that
\begin{equation} \label{split-inclusion}
   \mathcal{N}(O_1) \subset \mathcal{B} \subset \mathcal{N}(O_2) \,.
\end{equation}
A von Neumann algebra $\mathcal{B}$ is a {\it factor} if $\mathcal{B} \cap \mathcal{B}' = \mathbb{C}{\bf 1}$,
i.e.\ the intersection contains only multiples of the unit operator. The type I property means
that $\mathcal{B}$ is isomorphic, as a von Neumann algebra, to ${\sf B}(\hat{\mathcal{H}})$ for
some Hilbert space $\hat{\mathcal{H}}$. (Note the contrast to the generic type III property of the
local von Neumann algebras $\mathcal{N}_\omega(O)$ associated to spacetime regions $O$.) 
The split property implies normal product state extendability of normal states $\omega_A$ on $\mathcal{N}(O_A)$
and $\omega_B$ on $\mathcal{N}(O_B)$ as soon as $O_A \in O_1$ and $O_B \subset O_2^\perp$, and there is also
a form of a converse statement
\cite{DAnLon}. The split property has been shown to hold in several models of quantum field theory.
It can be deduced from a phase-space regularity condition in quantum field theory that guarantees 
stable thermodynamical behaviour \cite{BuWich}. 
Again, we refer to \cite{BuSplit,Verch-Nuc,Haag,DAnHol,FewsterSplit} and the reviews
\cite{Summers-Indep,Fewster-Verch-Review} and references given there for further discussion.

\section{The D-CTC condition in arbitrarily localizable QFT on a globally
hyperbolic spacetime}

\setcounter{equation}{0}

In this section, we will present our results on states that satisfy --- or do not satisfy ---
Deutsch's condition in the variants \eqref{DC} or \eqref{DCepsilon}, as discussed in the 
introduction. The results will be accompanied by remarks and comments 
on their interpretation.

The setting in which our presentation of results is staged is as follows.
As outlined in the introduction, we consider an arbitrarily localizable 
quantum field theory $\{ \mathcal{A}(O) \}_{O \in \mathcal{K}(\boldsymbol{M})}$
over a globally hyperbolic spacetime $\boldsymbol{M} = (M,g)$ of dimension
$d \ge 2$. The index set $\mathcal{K}(\boldsymbol{M})$ will be taken to consist
of all open, relatively compact subsets $O$ of $M$ (this means that the closure
of $O$ is a compact subset of $M$). In keeping with the previous discussion,
it will be assumed that the theory fulfills the conditions of isotony and locality,
and also additivity. 

First, we specialize the setting a bit more by assuming that the spacetime $\boldsymbol{M}$
is stationary, with time-shift isometry group $\{\vartheta_t\}_{t \in \mathbb{R}}$ and that the quantum field theory 
has the time-shift covariance property \eqref{timeshiftcovariance} with a 1-parametric group $\{\alpha_t\}_{t \in
\mathbb{R}}$ on $\mathcal{A}(\boldsymbol{M})$. Moreover, we assume the existence of a ground state $\omega_0$.
By $\mathcal{N}(O) = \pi_0(\mathcal{A}(O))''$ we denote the local von Neumann algebras in the GNS-representation
of $\omega_0$. We also choose two acausally spacetime regions $O_A$ and $O_B$ in $\mathcal{K}(\boldsymbol{M})$,
$O_A \subset O_B^\perp$. 

In this situation, we have:
\begin{proposition} \label{prop1}
Suppose that the timelike tube property \eqref{ttt-prop} holds in the GNS-representation of $\omega_0$.
Let $\tilde{\omega}({\bf a}) = {\rm Tr}(\tilde{\varrho} {\bf a})$ $({\bf a} \in {\sf B}(\mathcal{H}_0))$ be a normal state
in the GNS-representation of the ground state, given by a density matrix $\tilde{\varrho}$, and assume that
the state is analyic. Moreover, let $U \in {\sf B}(\mathcal{H}_0)$
be a unitary operator and assume that the state $\tilde{\omega}(U^\ast{\bf b}U) = {\rm Tr}(\tilde{\varrho} U^\ast {\bf b} U)$ 
$({\bf b} \in {\sf B}(\mathcal{H}_0))$ is also
analytic. If the two states coincide on $\mathcal{N}(O_B)$,
\begin{equation}
 \tilde{\omega}({\bf b}) = \tilde{\omega}(U^\ast {\bf b} U) \quad ( {\bf b} \in \mathcal{N}(O_B))\,,
\end{equation}
then the states $\tilde{\omega}$ and $\tilde{\omega}(U^\ast \,.\,U)$ coincide on ${\sf B}(\mathcal{H}_0)$. Equivalently,
$U\tilde{\varrho}U^\ast = \tilde{\varrho}$. 
\end{proposition}
{\it Proof}. The proof derives immediately from Lemma \ref{ttt-equality}. \hfill $\Box$
\\[6pt]
This may be seen as a particular form of a no-go result for a state $\tilde{\omega}$ that fulfills a very strong
form of entanglement and provides a solution to the problem of finding, for given unitary $U$, and given normal
state $\omega_A$ on $\mathcal{N}(O_A)$, a normal state $\tilde{\omega}$ so that
\begin{equation} \label{Deutschproblem}
 \tilde{\omega}({\bf a}) = \omega_A({\bf a}) \quad \text{and} \quad \tilde{\omega}(U^\ast {\bf b} U) = \tilde{\omega}({\bf b})
\end{equation}
holds for all ${\bf a} \in \mathcal{N}(O_A)$ and ${\bf b} \in \mathcal{N}(O_B)$. For not only does the state not change under
the action of $U$ on the ``System part $B$'', but it cannot change under action of $U$ on ``System part $A$'' either.
If $\tilde{\omega}$ could be obtained for a wide range of choices of $U$ and $\omega_A$, there would have to be changes
on ``System part $A$'' under the action of $U$
at least for some choices of $\omega_A$ and $U$. Under the stated assumptions, that is not possible.

One might object here that the assumptions do not only impose a very strong from of entanglement on the ``solution''
state $\tilde{\omega}$ to the problem \eqref{Deutschproblem} but also on the state $\tilde{\omega}(U^\ast \,. \,U)$.
One can give some motivation for this assumption, however. First, any unitary $U$ can be approximated, in the 
strong operator topology, by ${\rm e}^{iG_n}$ with a sequence of bounded symmetric operators $G_n \in {\sf B}(\mathcal{H}_0)$
$(n \in \mathbb{N})$. In turn,
any unitary operator in the sequence can be approximated, in the strong operator topology, by ${\rm e}^{iEG_nE}$ where the
$E$ denote spectral projections of the ground state Hamiltonian $H$ corresponding to finite spectral intervals.
Then, if $\tilde{\omega}$ is analytic, the same holds also for $\tilde{\omega}({\rm e}^{-iEG_nE}\, \,. \,\, {\rm e}^{iEG_nE})$,
and ${\rm e}^{iEG_nE}$ can be seen as a finite-energy cut-off approximation to the dynamical evolution $U$. That is something
which is quite legitimate to consider in place of the full dynamics in many problems.

However, one could try and keep the assumption of a strongly entangled $\tilde{\omega}$, but turn the assumptions 
on $\tilde{\omega}(U^\ast\,.\,U)$ into something qualitatively opposite to the case just considered. Nevertheless,
that does not improve the situation, as the next result shows.
Let us point out what we mean by opposite assumptions on $U$ compared to the previous case where we might envisage
$U$ as ${\rm e}^{iEG_n E}$. The occurrence of the spectral projections $E$ of the Hamiltonian corresponding to finite
energy intervals will prevent ${\rm e}^{iE G_n E}$ from being a local operator, i.e.\ from being contained in any
$\mathcal{N}(O)$, $O \in \mathcal{K}(\boldsymbol{M})$ (with the additional condition that $O^\perp$ is non-void in case that
$\boldsymbol{M}$ admits compact Cauchy surfaces). 
Thus, one assumption opposite to the previous case is to demand that $U$ is contained in a local von Neumann algebra.
Furthermore, since $E\psi$ (if non-zero) has the Reeh-Schlieder property even if 
$\psi$ induces a product state between local algebras corresponding to acausally related spacetime regions --- in the presence
of the split property, there are many such vectors $\psi$ --- the action of a unitary of the form ${\rm e}^{iE G_n E}$ on a 
state will typically change the correlations between observables located in acausally related spacetime regions drastically.
The opposite condition on $U$ is to require that such correlations in the state $\tilde{\omega}(U^\ast\,.\,U)$ 
are comparable to the correlations
in the state $\tilde{\omega}$, in a sense we will make precise in the following Def.\ \ref{compCorrel}.

In the following discussion, we don't need the assumption of a stationary spacetime or 
existence of a ground state, so now we assume only that $\{ \mathcal{A}(O)\}_{O \in \mathcal{K}(\boldsymbol{M})}$ is an arbitrarily
localizable quantum field theory on a globally hyperbolic spacetime $\boldsymbol{M}$. Furthermore, $\omega_0$ is now some
state in $\mathscr{S}(\boldsymbol{M})$ and again, $(\mathcal{H}_0,\pi_0,\Omega_0)$ denotes its GNS-representation and 
$\mathcal{N}(O) = \pi_0(\mathcal{A}(O))''$ are the local von Neumann algebras in that GNS representation.
\begin{definition} \label{compCorrel}
 Let three spacetime regions $O_1,O_2,O_3 \in \mathcal{K}(\boldsymbol{M})$ be given, where
 $O_1 \subset O_2$ and $O_3 \subset O_2^\perp$. Let $\tilde{\omega}$ be a normal state on ${\sf B}(\mathcal{H}_0)$
 and let $U \in \mathcal{N}(O_2)$. Furthermore, suppose that
 \begin{equation} \label{omegaUinvar1}
  \tilde{\omega}(U^\ast {\bf a}_1 U) = \tilde{\omega}({\bf a}_1)
 \end{equation}
 for all ${\bf a}_1 \in \mathcal{N}(O_1)$.
 Then we say that $\tilde{\omega}$ and $\tilde{\omega}(U^\ast\,.\,U)$
 have {\em comparable correlations between $\mathcal{N}(O_1)$ and $\mathcal{N}(O_3)$} if there is a 
 number $q \ge 0$ such that the estimates 
 \begin{equation} \label{EstCompCor}
 \begin{split}
  | \,\tilde{\omega}(U^\ast {\bf a}_1^\ast {\bf a}_3 U) - (q + 1) \tilde{\omega}({\bf a}_1^\ast {\bf a}_3)\, | 
  & \le \frac{q}{2} \left(\tilde{\omega}({\bf a}_1^\ast{\bf a}_1) + \tilde{\omega}({\bf a}_3^\ast{\bf a}_3)\right) \,,\\
 |\, \tilde{\omega}( {\bf a}_1^\ast {\bf a}_3) - (q + 1) \tilde{\omega}(U^*{\bf a}_1^\ast {\bf a}_3U) \, | 
  & \le \frac{q}{2} \left(\tilde{\omega}({\bf a}_1^\ast{\bf a}_1) + \tilde{\omega}({\bf a}_3^\ast{\bf a}_3)\right)  
 \end{split}
 \end{equation}
hold for all ${\bf a}_1 \in \mathcal{N}(O_1)$ and ${\bf a}_3 \in \mathcal{N}(O_3)$. 
\end{definition}
The positive parameter $q$ serves as a measure for the deviation of the correlations in the states
$\tilde{\omega}$ and $\tilde{\omega}(U^\ast\,.\,U)$. The deviation becomes larger with increasing $q$, and
vanishes for $q = 0$. 

Let us next assume that $\tilde{\omega}(\,.\,) = \langle \tilde{\psi},\,.\,\tilde{\psi}\rangle$ is
induced by a unit vector $\tilde{\psi} \in \mathcal{H}_0$. With the notation as in Def.\ \ref{compCorrel},
one obtains:
\begin{lemma} \label{correlUdominance}
 The states $\tilde{\omega}$ and $\tilde{\omega}(U^\ast\,.\,U)$
 have comparable correlations between $\mathcal{N}(O_1)$ and $\mathcal{N}(O_3)$ if and only if
 there is a number $K \ge 1$ such that the estimates
 \begin{equation}
  || ({\bf a_1} + {\bf a}_3)U\tilde{\psi}|| \le K||({\bf a_1} + {\bf a}_3)\tilde{\psi}|| \quad 
  \text{and} \quad 
  || ({\bf a_1} + {\bf a}_3)\tilde{\psi}|| \le K||({\bf a_1} + {\bf a}_3)U\tilde{\psi}||
  \end{equation}
hold for all ${\bf a}_1 \in \mathcal{N}(O_1)$ and ${\bf a}_3 \in \mathcal{N}(O_3)$.  
\end{lemma}
{\it Proof.} 
Note that the assumptions on $\tilde{\omega}$, $U$ and the spacetime regions $O_1,O_2,O_3$ imply
that $U \in \mathcal{N}(O_3)'$ and so we have
\begin{equation} \label{omegaUinvar3}
  \tilde{\omega}(U^\ast {\bf a}_3 U) = \tilde{\omega}({\bf a}_3)
 \end{equation}
 for all ${\bf a}_3 \in \mathcal{N}(O_3)$.
 
Now suppose that $\tilde{\omega}$ and $\tilde{\omega}(U^\ast\,.\,U)$
 have comparable correlations between $\mathcal{N}(O_1)$ and $\mathcal{N}(O_3)$.
The first estimate of \eqref{EstCompCor} implies
\begin{equation}
2{\rm Re}\,( \,\tilde{\omega}(U^\ast {\bf a}_1^\ast {\bf a}_3 U) - (q + 1) \tilde{\omega}({\bf a}_1^\ast {\bf a}_3)) \,) 
  \le {q} \left(\tilde{\omega}({\bf a}_1^\ast{\bf a}_1) + \tilde{\omega}({\bf a}_3^\ast{\bf a}_3)\right)\,.
\end{equation}
Setting $K^2 = q +1$, the previous estimate is equivalent to
\begin{equation}
\tilde{\omega}({\bf a}_1^\ast{\bf a}_1) + \tilde{\omega}({\bf a}_3^\ast{\bf a}_3)
+ 2{\rm Re}\,\tilde{\omega}(U^\ast {\bf a}_1^\ast {\bf a}_3 U) 
\le K^2 \left(\tilde{\omega}({\bf a}_1^\ast{\bf a}_1) + \tilde{\omega}({\bf a}_3^\ast{\bf a}_3)
+ 2{\rm Re}\,\tilde{\omega}({\bf a}_1^\ast {\bf a}_3 )\right)\,.
\end{equation}
Observing \eqref{omegaUinvar1} and \eqref{omegaUinvar3}, this is equivalent to
\begin{equation}
  || ({\bf a_1} + {\bf a}_3)U\tilde{\psi}|| \le K||({\bf a_1} + {\bf a}_3)\tilde{\psi}||\,.
\end{equation}
The other implications are obtained by similar arguments.
\hfill $\Box$
\\ \\
\begin{proposition} \label{prop2}
 Suppose that $\tilde{\psi}$ is a unit vector in $\mathcal{H}_0$ so that
 the state  $\tilde{\omega}(\,.\,) = \langle \tilde{\psi},\,.\,\tilde{\psi} \rangle$ 
 fufills \eqref{Deutschproblem} for some given normal state $\omega_A$ on $\mathcal{N}(O_A)$
 and unitary $U \in {\sf B}(\mathcal{H}_0)$. 
 
 Then the assumptions
 \begin{itemize}
  \item[(I)] $\tilde{\psi}$ has the Reeh-Schlieder property,
  \item[(II)] $U \in \mathcal{N}(O_U)$ for some $O_U \in \mathcal{K}(\boldsymbol{M})$ with
  $O_A \cup O_B \subset O_U$, 
  \item[(III)] there is some $O_C \in \mathcal{K}(\boldsymbol{M})$ with $\overline{O_C} \subset O_U^\perp$
  so that $\tilde{\omega}$ and $\tilde{\omega}(U^\ast \,.\, U)$ have comparable correlations
  between $\mathcal{N}(O_B)$ and $\mathcal{N}(O_C)$
  \end{itemize}
 together
imply that $U \in \mathcal{N}(O_B)'$\,.

\end{proposition}
We remark that there is a dense set of states (in the set of states fulfilling the microlocal spectrum condition) 
obeying the Reeh-Schlieder property
for quantized fields subject to linear field equations also in the case that the spacetime is not
stationary, see \cite{Verch-AntilocRS,DAnHol,Sanders-ReehSchl}.

Furthermore, the assumption that $\tilde{\omega}$ is induced by a unit vector $\tilde{\psi}$ may
appear restrictive, but in fact, this is not the case. For one could actually work in the GNS
representation of $\tilde{\omega}$ where the state $\tilde{\omega}$ is induced by the GNS vector
$\Omega_{\tilde{\omega}}$. On the other hand, even if one does not work in the GNS representation
of $\tilde{\omega}$, only properties of $\tilde{\omega}$ on the von Neumann 
algebra $(\mathcal{N}(O_U) \cup \mathcal{N}(O_C))''$ are tested. Under very general conditions, this
von Neumann algebra will be of type ${\rm III}_1$ and any normal state on this algebra is 
hence induced by a unit vector in the underlying Hilbert space, as indicated at the end of Sec.\
\ref{AnRSttt}.
\\[6pt]
{\it Proof.} 
The first observation is that in view of Lemma \ref{correlUdominance} 
there is a constant $K > 1$ with 
\begin{equation} \label{Uestimate2}
  || ({\bf b} + {\bf c})U\tilde{\psi}|| \le K ||({\bf b} + {\bf c})\tilde{\psi}|| \quad \text{and}\quad
  || ({\bf b} + {\bf c})\tilde{\psi}|| \le K ||({\bf b} + {\bf c})U\tilde{\psi}||
\end{equation}
for all ${\bf b} \in \mathcal{N}(O_B)$ and ${\bf c} \in \mathcal{N}(O_C)$. We define the linear
operator 
\begin{equation}
 W: ({\bf b} + {\bf c})\tilde{\psi} \mapsto ({\bf b} + {\bf c})U\tilde{\psi} \quad 
 ({\bf b} \in \mathcal{N}(O_B),\ {\bf c} \in \mathcal{N}(O_C)) \,.
\end{equation}
Note that the operator is well-defined: By having assumed $\overline{O_C} \subset O_U^\perp$, 
there is some local von Neumann algebra $\mathcal{N}(O') \subset (\mathcal{N}(O_B) \cup \mathcal{N}(O_C))'$.
The Reeh-Schlieder property of $\tilde{\psi}$ then implies that $\tilde{\psi}$ is cyclic and separating
for $(\mathcal{N}(O_B) \cup \mathcal{N}(O_C))''$. Consequently, ${\bf b} + {\bf c}$ is uniquely determined by
$({\bf b} + {\bf c})\tilde{\psi}$. Furthermore, the range of $W$ is dense: Setting ${\bf b} = 0$ one has
$W({\bf c}\tilde{\psi}) = {\bf c}U\tilde{\psi} = U{\bf c}\tilde{\psi}$; by the Reeh-Schlieder property,
$\mathcal{N}(O_C)\tilde{\psi}$ is dense in $\mathcal{H}_0$ and as $U$ is unitary, the set of 
all $U{\bf c}\tilde{\psi}$, ${\bf c} \in \mathcal{N}(O_C)$ is dense in $\mathcal{H}_0$. 
Linearity of $W$ is obvious. Furthermore, by \eqref{Uestimate2} $W$ extends to a bounded operator
on $\mathcal{H}_0$, and again by \eqref{Uestimate2}, we conclude that $W$ has a bounded inverse 
operator $W^{-1}$.

From the definition of $W$ we have
$W{\bf c}\tilde{\psi} = {\bf c}U\tilde{\psi} = U{\bf c}\tilde{\psi}$ for all ${\bf c} \in \mathcal{N}(O_C)$,
hence $W = U$. On the other hand,
we also have for all ${\bf b} \in \mathcal{N}(O_B)$ and all ${\bf c} \in \mathcal{N}(O_C)$,
\begin{equation}
W{\bf b}W^{-1}{\bf c}U\tilde{\psi} = W{\bf b}W^{-1}W{\bf c}\tilde{\psi} = W{\bf b}{\bf c}\tilde{\psi} = {\bf c}W{\bf b}\tilde{\psi}
= {\bf c}{\bf b}U\tilde{\psi} = {\bf b}{\bf c}U\tilde{\psi}
\end{equation}
where we have used that $W = U \in \mathcal{N}(O_C)'$. Using that $\mathcal{N}(O_C)U\tilde{\psi}$ is dense in $\mathcal{H}_0$,
we conclude that $W = U \in \mathcal{N}(O_B)'$. 
\hfill $\Box$
\\[8pt]
Again, this yields a negative result for the problem of finding, for given unitary $U$ and 
normal state $\omega_A$, a normal state $\tilde{\omega}$ satisfying \eqref{Deutschproblem} insofar as under
the given assumptions, $U$ must commute with all ${\bf b} \in \mathcal{N}(O_B)$, implying that 
$U$ does not act on ``System part $B$'', and therefore does not induce any dynamical coupling between
``System part $A$'' and ``System part $B$''. The requirement $\tilde{\omega}(U^\ast {\bf b}U) = \tilde{\omega}({\bf b})$
for all ${\bf b} \in \mathcal{N}(O_B)$ is hence trivially fulfilled. 

These results indicate that the Reeh-Schlieder property might be an obstacle for a state $\tilde{\omega}$
to induce a solution to the Deutsch condition problem \eqref{Deutschproblem}. However, if one drops the 
requirement that $\tilde{\omega}$ fulfill the Reeh-Schlieder property, matters look much different, as we shall see
next.

We consider an arbitrarily localizable quantum field theory on a globally hyperbolic spacetime
$\boldsymbol{M}$ with family of local von Neumann algebras $\{\mathcal{N}(O)\}_{O \in \mathcal{K}(\boldsymbol{M})}$
in the GNS representation $(\mathcal{H},\pi\Omega)$ of a state $\omega \in \mathscr{S}(\boldsymbol{M})$. 
Moreover, we set $\mathcal{N}_A = \mathcal{N}(O_A)$ and $\mathcal{N}_B = \mathcal{N}(O_B)$ for 
a pair of $O_A,O_B \in \mathcal{K}(\boldsymbol{M})$ that are properly acausally related.
For the next result, we assume that for any pair of normal states $\varphi_A$ on $\mathcal{N}_A$ and $\varphi_B$ on
$\mathcal{N}_B$ there is a normal state on ${\sf B}(\mathcal{H})$ which extends $\varphi_A$ and $\varphi_B$, so that
${\rm NEx}(\varphi_A,\varphi_B)$ is non-empty. As mentioned in the previous section, a sufficient condition for this to hold is that 
the local von Neumann algebras $\mathcal{N}(O)$, $O \in \mathcal{K}(\boldsymbol{M})$,
fulfill the split property which has been shown to hold in several quantum field theoretical models.
\begin{proposition} \label{approxiDeutschinQFT}
Suppose 
that ${\rm NEx}(\varphi_A,\varphi_B)$ is non-empty for any pair of normal states $\varphi_A$ on $\mathcal{N}_A$
and $\varphi_B$ on $\mathcal{N}_B$.
 Let a normal state $\omega_A$ on $\mathcal{N}_A$ and a unitary
operator $U$ in ${\sf B}(\mathcal{H})$ be given, as well as any positive numbers $\varepsilon$ and $R$.
Then there is a normal state $\tilde{\omega}$ on
${\sf B}(\mathcal{H})$  such that 
\begin{equation}
 \tilde{\omega}({\bf a}) = \omega_A({\bf a}) \quad \text{and} \quad 
 |\, \tilde{\omega}(U^\ast {\bf b} U) - \tilde{\omega}({\bf b}) \,| < \varepsilon
\end{equation}
holds for all ${\bf a} \in \mathcal{N}_A$ and all ${\bf b} \in \mathcal{N}_B$ with $||{\bf b}|| \le R$.
\end{proposition}
{\it Proof.} \ \ For any normal state $\varphi$ on ${\sf B}(\mathcal{H})$, let us write
\begin{equation}
 u[\varphi]({\bf c}) = \varphi(U^\ast {\bf c} U) \quad ({\bf c} \in {\sf B}(\mathcal{H}))\,.
\end{equation}
Moreover, the restriction of $\varphi$ to $\mathcal{N}_A$ or $\mathcal{N}_B$
will be denoted by ${\tt r}_A\varphi$, respectively ${\tt r}_B\varphi$, so that
\begin{equation}
     {\tt r}_A\varphi({\bf a}) = \varphi({\bf a}) \quad \text{and} \quad {\tt r}_B\varphi({\bf b}) = \varphi({\bf b}) 
\end{equation}
for all ${\bf a} \in \mathcal{N}_A$ and all ${\bf b} \in \mathcal{N}_B$. 

By assumption,  ${\rm NEx}(\varphi_A,\varphi_B)$ is non-empty
whenever $\varphi_A$ is a normal state on $\mathcal{N}_A$ and $\varphi_B$ is normal state
on $\mathcal{N}_B$. We will use this fact to establish, in the first step of our proof, the
existence of a (non-unique) sequence $\{\varphi_n\}_{n \in \mathbb{N}}$ of normal states on ${\sf B}(\mathcal{H})$
with the properties that
\begin{equation} \label{iterative}
 {\tt r}_B \varphi_{n} = {\tt r}_B u[\varphi_{n - 1}] \ \ (n \in \mathbb{N}, n \ge 2) \ \ \text{and} \ \
  {\tt r}_A\varphi_{n} = \omega_A \ \ (n \in \mathbb{N})\,.
\end{equation}
To establish this, choose any normal state $\omega_B$ on $\mathcal{N}_B$, and choose as $\varphi_1$
any state in ${\rm NEx}(\omega_A,\omega_B)$. Obviously, it holds that
${\tt r}_A\varphi_1 = \omega_A$. Next, we choose a state $\varphi_2$ in
${\rm NEx}(\omega_A,{\tt r}_B(u[\varphi_1]))$. By construction, it holds that 
${\tt r}_A\varphi_2 = \omega_A$ and ${\tt r}_B\varphi_2 = {\tt r}_B(u[\varphi_1])$. 

The existence of a sequence $\{\varphi_n\}_{n \in \mathbb{N}}$ with the properties of
\eqref{iterative}
will then be concluded inductively. Therefore, let us pick any $k \ge 2$ in $\mathbb{N}$ and suppose 
that states $\varphi_1,\ldots,\varphi_k$  with the required properties \eqref{iterative} (for $n \le k$) have been
selected. We pick some $\varphi_{k +1} \in {\rm NEx}(\omega_A,{\tt r}_B(u[\varphi_k]))$ and
we claim that the resulting sequence $\varphi_1,\ldots,\varphi_k,\varphi_{k+1}$ has the 
properties as in \eqref{iterative} (for $n \le k + 1$) as well. However, the points we need to check, namely
${\tt r}_B\varphi_{k+1} = {\tt r}_B(u[\varphi_k])$ and ${\tt r}_A\varphi_{k +1} = \omega_A$
result simply from  $\varphi_{k +1} \in {\rm NEx}(\omega_A,{\tt r}_B(u[\varphi_k]))$.

In the next step, we construct a state $\tilde{\omega}$ having the properties as stated in the
proposition from a sequence $\{\varphi_n\}_{n \in \mathbb{N}}$ 
of normal states subject to \eqref{iterative}. 
For any $N \in \mathbb{N}$, let 
\begin{equation}
 \tilde{\omega}_N = \frac{1}{N} \sum_{n = 1}^N \varphi_n  \,.
\end{equation}
Then it holds that
\begin{equation}
  {\tt r}_A(\tilde{\omega}_N) = \ \frac{1}{N} \sum_{n = 1}^N {\tt r}_A  \varphi_n 
      = \frac{1}{N} \sum_{n =1}^N \omega_A 
      =  \omega_A\,;
\end{equation}
equivalently, $\tilde{\omega}_N({\bf a}) = \omega_A({\bf a})$ $({\bf a} \in \mathcal{N}_A)$ for all $N \in \mathbb{N}$.
Moreover, for any ${\bf b} \in \mathcal{N}_B$ one finds 
\begin{equation}
\begin{split}
  | u[\tilde{\omega}_N]({\bf b}) - \tilde{\omega}_N({\bf b})| 
   & = \ \left| \frac{1}{N} \left( \sum_{n = 1}^N u[\varphi_n]({\bf b}) - \varphi_n({\bf b}) \right) \right| \\
   & = \ \frac{1}{N} \left| u[\varphi_N]({\bf b}) - \varphi_1({\bf b}) \right| \\
\end{split}
\end{equation}
where we have passed to the last equality on account of \eqref{iterative}.
Thus, choosing $N$ larger than $2R/\varepsilon$,
one obtains for all ${\bf b} \in \mathcal{N}_B$ with $||{\bf b}|| \le R$ the bound
\begin{equation}
 | \tilde{\omega}_N(U^\ast {\bf b}U) - \tilde{\omega}_N({\bf b})| < \varepsilon \,.
\end{equation}
Then setting $\tilde{\omega} = \tilde{\omega}_N$ concludes the proof. ${}$ \hfill $\Box$
\\[10pt]
The proof follows the same pattern as the analogous proof of \eqref{dc-trversion} in \cite{Deutsch}
for finite-dimensional $\mathcal{H}_A$ and $\mathcal{H}_B$, adapted to the more general, infinite-dimensional
situation. We emphasize that we require the state $\tilde{\omega}$ to be a {\it normal} state, i.e.\ given
by a density matrix in the GNS representation of the initially chosen state $\omega$. This requirement
is the reason why --- with this method of proof --- the invariance of $\tilde{\omega}$ under the action
of $U$ on $\mathcal{N}_B$ can only be obtained approximately. It is still possible that in some cases the sequence of 
normal states $\tilde{\omega}_N$ might converge as $N \to \infty$ and then the limiting state $\tilde{\omega}$
will indeed satisfy \eqref{Deutschproblem} exactly. However, in general it is not known if the $\tilde{\omega}_N$
will converge as $N \to \infty$. Still, the sequence of normal states $\tilde{\omega}_N$ possesses limit points, as
$N \to \infty$, in the weak $C^*$ sense by the Banach-Alaoglu theorem \cite{ReedSimon1}, but as known from examples, such
limit points will often not be normal states in the underlying Hilbert space representation anymore. We provide 
an example in the Appendix. If one were to drop the requirement of $\tilde{\omega}$ to be normal, the 
limit point states provide exact solutions to \eqref{Deutschproblem} as states on the $C^*$ algebra ${\sf B}(\mathcal{H})$,
but in case they are not normal, they have unwieldy continuity properties with respect to convergence of vectors
in $\mathcal{H}$, and are not interpretable as physically realistic states. Nevertheless, the result shows that under
the given assumptions, the D-CTC condition \eqref{Deutschproblem} always has approximate solutions $\tilde{\omega}$ to arbitrary
accuracy in quantum field theory on any globally hyperbolic spacetime.

Finally we note that the state $\tilde{\omega}$ of Prop.\ \ref{approxiDeutschinQFT} (contructed as $\tilde{\omega}_N$ in the proof)
is not unique. It should also be noted that in the case where
the quantum field theory satisfies the split property, the states $\varphi_n$ can be chosen as normal product states
with respect to $\mathcal{N}_A$ and $\mathcal{N}_B$, and then
the state $\tilde{\omega}$ (constructed as one of the $\tilde{\omega}_N$ for sufficiently large $N$) 
is a convex sum of product states across $\mathcal{N}_A$ and $\mathcal{N}_B$ and consequently shows no quantum
entanglement between ``System part $A$'' and ``System part $B$''.

\section{The quantized massless Klein-Gordon field on the Politzer spacetime}  \setcounter{equation}{0}

As has been mentioned in the Introduction, the Politzer spacetime has been suggested as a spacetime analogue interpretation
of situations to which the D-CTC criterion applies, and where there is a spacetime part with a causality preserving dynamical
evolution coupled to a dynamical evolution in a spacetime part with CTCs. 

Here, we would like to discuss further some aspects of quantum field theory on that particular spacetime, restricting
to the case of the free real, massless field.

To begin, one is confronted with the difficulty that the Politzer spacetime is not in a natural way a Lorentzian $C^\infty$
spacetime without further specification, in particular, of a $C^\infty$ atlas near the boundary points of the cut-away strips
$S_\pm$ (see Fig.\ 2). There is a way to circumvent this in the case where one is interested in (quantized) fields obeying 
a linear hyperbolic partial differential equation
brought into the form
\begin{equation} \label{anotherDeqn}
  \partial_t \psi(t,x) = D_x\psi(t,x)
\end{equation}
as pointed out in the introduction, with $\psi(t,x) \in \mathbb{C}^N$ and $D_x$ a suitable (matrix-valued) differential operator
with respect to $x$. In this case, one can start by defining $\psi$ on Minkowski spacetime where the strips $S_\pm$ have been
cut away, and impose boundary conditions at the ``upper'' and ``lower'' rims of the strips as in \eqref{initdataconstraint} to
model the effect of CTCs inbetween the strips.

For the sake of simplicity, we will here consider in place of a general linear hyperbolic field $\psi$ the 
{\it real, massless Klein-Gordon field}. On two-dimensional Minkowski spacetime $M = \mathbb{R}^{1,1}$ with
inertial time-coordinate $t$ and space-coordinate $x$, the (classical) real, massless Klein-Gordon field stands for any solution
$\phi_M \in C^\infty(M,\mathbb{R})$ of
\begin{equation}
 \Box\phi_M(t,x) = (\partial^2_t - \partial_x^2)\phi_M(t,x) = 0 \,.
\end{equation}
We are interested in the subspace of those solutions which have {\it compactly supported Cauchy data} and denote that by 
$\mathcal{S}_M$. As is well known, one can introduce a symplectic form $\sigma_M$ on 
$\mathcal{S}_M$ by 
\begin{equation}
  \sigma_M(\phi_M,\tilde{\phi}_M) = \int_{ \Sigma_0 } \left( (\partial_0 \phi_M)\tilde{\phi}_M - \phi_M \partial_0 \tilde{\phi}_M \right) \, dx \quad
  (\phi_M,\tilde{\phi}_M \in \mathcal{S}_M)
\end{equation}
where $\Sigma_0$ is the $t = 0$ Cauchy-surface and $\partial_0$ stands for $\partial_t |_{t = 0}$. One can replace $\Sigma_0$ by any 
other Cauchy-surface and $\partial_0$ by the corresponding future-pointing normal derivative without changing the value 
of $\sigma_M(\phi_M,\tilde{\phi}_M)$. 

As is also well-known, any $\phi_M$ in $S_M$ can be written uniquely as the sum of a ``right-moving'' part $\phi_M^R$ and a 
``left-moving'' part $\phi^L_M$ which are both in $\mathcal{S}_M$ and characterized by the property that there are 
functions $\xi_R$ and $\xi_L$ in $C_0^\infty(\mathbb{R},\mathbb{R})$ so that\footnote{In order to avoid 
infrared problems on constructing a vacuum representation for the quantized field, it is customary to 
restrict the $\xi_R$ and $\xi_L$ to be derivatives of $C_0^\infty(\mathbb{R},\mathbb{R})$ functions. See the section
on the ``Schwinger model'' in
\cite{BLTO} for discussion.}
\begin{equation}
 \phi_M^R(t,x) = \xi_R(t - x) \,, \quad \ \ \phi_M^L(t,x) = \xi_L(t + x) \quad \ \ \ ( (t,x) \in M)\,.
\end{equation}
Furthermore, any right- and left-moving elements in $\mathcal{S}_M$ are symplectically orthogonal,
\begin{equation}
 \sigma_M(\phi_M^R,\tilde{\phi}_M^L) = 0 \quad (\phi_M,\tilde{\phi}_M \in \mathcal{S}_M)\,.
\end{equation}

A $C^*$-algebraic quantization of the real, massless Klein-Gordon field on 2-dimensional Minkowski spacetime can now be 
obtained in a completely standard manner (cf.\ e.g. \cite{KayGround}): One defines the {\it Weyl algebra} 
${\sf W}(\mathcal{S}_M,\sigma_M)$ corresponding 
to the symplectic space $(\mathcal{S}_M,\sigma_M)$; this is the unique $C^*$-algebra generated by a unit element ${\bf 1}$ and 
elements ${\bf w}(\phi_M)$, $\phi_M \in \mathcal{S}_M$, subject to the relations
\begin{equation}
 {\bf w}(\phi_M)^* = {\bf w}(-\phi_M)\,, \quad {\bf w}(0) = {\bf 1}\,, \quad {\bf w}(\phi_M){\bf w}(\tilde{\phi}_M) = 
  {\rm e}^{i\sigma_M(\phi_M,\tilde{\phi}_M)/2}{\bf w}(\phi_M + \tilde{\phi}_M) \,.
\end{equation}
Then $\mathcal{A}(M) = {\sf W}(\mathcal{S}_M,\sigma_M)$ will constitute the $C^*$-algebra of observables of 
the quantized real, massless Klein-Gordon field on 2-dimensional Minkowski spacetime.
Similarly one can introduce the sub-$C^*$-algebras $\mathcal{A}^R(M)$ and $\mathcal{A}^L(M)$ which are generated by all 
${\bf w}(\phi_M^R)$ and ${\bf w}(\phi^L_M)$ $(\phi_M \in \mathcal{S}_M)$, respectively. Clearly, $\mathcal{A}^R(M)$
and $\mathcal{A}^R(M)$ are commuting subalgebras of $\mathcal{A}(M)$, and together they generate $\mathcal{A}(M)$. 
 
Given an open, relatively compact subset $O$ of $M$, one defines the local $C^*$-algebra $\mathcal{A}_M(O)$ associated with
the localization region $O$ as the $C^*$-subalgebra of $\mathcal{A}(M)$ generated by all ${\bf w}(\phi_M)$ with the property
that there is a
Cauchy surface $\Sigma$ in $M$ such that $\phi_M$ has compactly supported Cauchy-data in $O \cap \Sigma$. 
The system of local $C^*$-algebras $\{ \mathcal{A}_M(O) \}_{O \in \mathcal{K}(\mathbb{R}^{1,1})}$ can then be shown to
fulfill the conditions of locality and isotony. Furthermore, it carries a covariant action of the proper, orthochronous
conformal group on $M$. Here, we are only interested in the translations, and the action of the translation group
by automorphisms of $\mathcal{A}(M)$ is given by 
\begin{equation}
 \alpha_{(t',x')}({\bf w}(\phi_M)) = {\bf w}(\phi_{M,(t',x')}) \,, 
 \quad \phi_{M,(t',x')}(t,x) = \phi_M(t-t',x - x') \quad ((t',x') \in \mathbb{R}^2)
\end{equation}
yielding 
\begin{equation}
 \alpha_{(t',x')}(\mathcal{A}_M(O)) = \mathcal{A}_M(O + (t',x'))\,.
\end{equation}
The properties of the left- and right-moving elements in $\mathcal{S}_M$ imply
\begin{equation}
 \alpha_{(s,s)}({\bf w}(\phi_M^R))  = {\bf w}(\phi_M^R)\,, \quad \alpha_{(s,-s)}({\bf w}(\phi_M^L)) 
 = {\bf w}(\phi_M^L) \quad (\phi_M \in \mathcal{S}_M) 
\end{equation}
for all $s \in \mathbb{R}$ and hence,
\begin{equation}
 \alpha_{(s,s)}(\mathcal{A}^R_M(O)) = \mathcal{A}^R_M(O) \,, \quad \alpha_{(s,-s)}(\mathcal{A}^L_M(O)) = \mathcal{A}^L_M(O)
\end{equation}
for all $s \in \mathbb{R}$ and all relatively compact open subsets $O$ of $M$ with the obvious definition
\begin{equation}
  \mathcal{A}_M^R(O) = \mathcal{A}_M(O) \cap \mathcal{A}^R(M) \,, \quad \mathcal{A}^L_M(O) = \mathcal{A}_M(O) \cap \mathcal{A}^L(M)\,.
\end{equation}
Then the algebras $\mathcal{A}^R_M(O)$ and $\mathcal{A}^R_M(O)$ are commuting $C^*$-subalgebras of $\mathcal{A}_M(O)$, and 
together they generate $\mathcal{A}_M(O)$.

We note that additivity of the local algebras $\mathcal{A}_M(O)$ can also be shown to be fulfilled.
Furthermore, the {\it time-slice property} holds, i.e.\ $\mathcal{A}_M(D(O)) = \mathcal{A}_M(O)$ where we recall
that $D(O)$ denotes the domain of dependence of $O$. Hence, for any open, relatively compact subset $C$ of some
Cauchy-surface $C$ one can define $\mathcal{A}_M(C) = \mathcal{A}_M(D(C))$. The definition can be extended 
to infinitely extended $C$, and the time-slice property then implies that $\mathcal{A}_M(\Sigma) = \mathcal{A}(M)$ for
every Cauchy-surface $\Sigma$.

Now we would like to construct a theory of the quantized real, massless Klein-Gordon field on the Politzer spacetime.
As already mentioned, the difficulty of having to define a proper $C^\infty$ manifold structure for that spacetime
will be circumvented by defining the Klein-Gordon field on Minkowski spacetime where the strips $S_\pm$ have been
removed, and imposing appropriate boundary conditions. In more detail, we define
\begin{equation}
 P = \mathbb{R}^{1,1} \backslash ( S_+ \cup S_-)
\end{equation}
and we seek solutions $\phi_P \in C^\infty(P,\mathbb{R})$ to the partial differential equation
\begin{equation} \label{condKG}
 \Box\phi_P(t,x) = 0 \quad ((t,x) \in P)
\end{equation}
subject to the boundary conditions
\begin{equation}
\begin{split} \label{condPM}
 \lim_{\varepsilon \to 0+}\,\phi_P(\tau -\varepsilon,x) & = \lim_{\varepsilon \to 0+}\, \phi_P(-\tau + \varepsilon,x) \quad 
 (|x| \le L) \\
 \lim_{\varepsilon \to 0+}\,\phi_P(-\tau -\varepsilon,x) & = \lim_{\varepsilon \to 0+}\, \phi_P(\tau + \varepsilon,x) \quad 
 (|x| \le L) \,.
\end{split}
\end{equation}
(Recall the definition of $S_\pm$ in terms of the positive numbers $\tau$ and $L$ given in the Introduction.) 

These boundary conditions are ``minimal'' in order to capture the 
idea that the spacetime inbetween the removed strips contains CTCs, but
presumably they are not sufficient in order to allow statements on existence or uniqueness of solutions and dependence on
suitably posed intial data. What is missing are further boundary conditions at the endpoints of the removed
strips, i.e.\ at the points $(\tau,\pm L)$ and $(-\tau,\pm L)$. There may be various possibilities to endow
the Politzer spacetime with $C^\infty$ structures, and this then results in different possibilities for the 
behaviour of solutions $\phi_P$ at the endpoints of the $S_\pm$. It is not a priori excluded to admit singular behaviour 
of the solutions $\phi_P$ at these critical points, even though this might imply that the solutions are also
singular on other subsets of $P$. Moreover, the question if the endpoints of $S_\pm$ should be regarded as belonging
to the Politzer spacetime or not could also be settled in terms of boundary conditions for the solutions
$\phi_P$ at the endpoints of the $S_\pm$.

At this point, we cannot go into any further detail on this circle of problems, and shall be content with 
imposing a very restrictive additional condition at the boundary points of the $S_\pm$,
\begin{equation} \label{cond-0}
 \text{for every}\ \phi_P \ \text{there are open neighbourhoods of} \ (\tau,\pm L)\ \text{and}
 \ (-\tau,\pm L) \ \text{on which} \ \phi_P = 0 \,.
\end{equation}
This condition allows it to construct algebras of local observables $\mathcal{A}_P(O)$ for a quantized real, 
massless Klein-Gordon field
on the Politzer spacetime as $C^*$-subalgebras of  $\mathcal{A}(M)$ where the assignment of localization
regions $O$ to the $\mathcal{A}_P(O)$ differs from the assignment of localization regions $O$ to the local 
algebras $\mathcal{A}_M(O)$ on Minkowski spacetime. This just goes to illustrate the fundamental insight of 
local quantum physics, put forward by Haag and Kastler, that the core of physical information lies in 
the assignment of spacetime regions $O$ to local algebras $\mathcal{A}(O)$ \cite{HaagKastler,Haag}. 

Let us define by $\mathcal{S}_P^0$ the set of all solutions $\phi_P$ to \eqref{condKG} in $C^\infty(P,\mathbb{R})$
which satisfy the boundary conditions \eqref{condPM} and \eqref{cond-0}. Again, any solution $\phi_P$ in
$\mathcal{S}_P^0$ can be uniquely split into a right-moving part $\phi_P^R$ and a left-moving part $\phi_P^L$.
However, the boundary conditions \eqref{condPM} now impose continuation conditions on these solutions. In
particular, in the domain 
\begin{equation}
 P_{\rm CTC} = \{ (t,x) : -\tau < t < \tau\,, \ \ -L \le x \le L \}
\end{equation}
this leads to a form of periodicity of the propagation behaviour, as depicted in Figure 4.
\begin{center}
 \includegraphics[width=13cm]{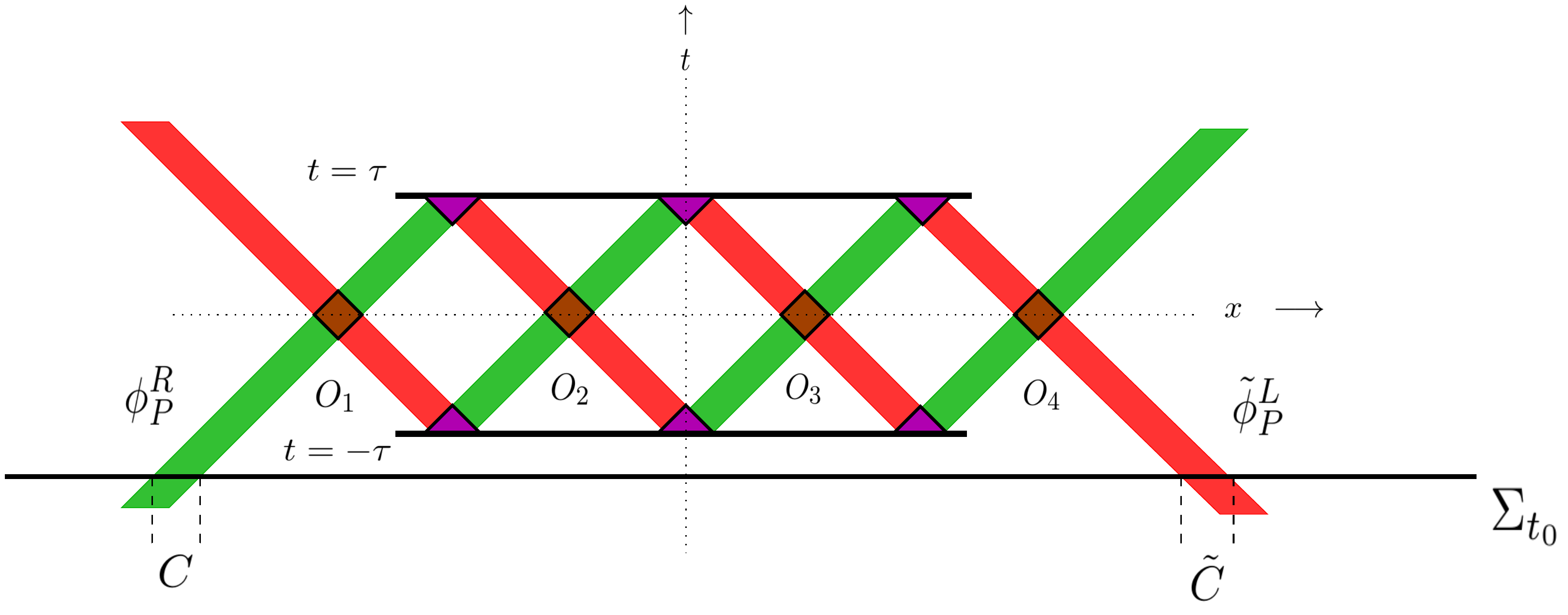}
\end{center}
{\small
{\bf Figure 4}. Supports of a right-moving solution $\phi^R_P$ (green) and a left-moving solution $\tilde{\phi}^L_P$
(red) of the real, massless Klein-Gordon equation entering the CTC-region $P_{\rm CTC}$. Their supports intersect
in the diamond-shaped regions (brown) denoted by $O_1,\ldots,O_4$, and also in the triangular regions (purple). 
The support regions $C$ and $\tilde{C}$ of the initial data on the $t = t_0$ hyperplane $\Sigma_{t_0}$ have been chosen
small enough so that any right/left moving solution emanating from these regions lies in $\mathcal{S}_P^0$ without
further constraints on the Cauchy data.
}
\\[8pt]
A way to construct local algebras of observables $\mathcal{A}_P(O)$ for the quantized real,
massless Klein-Gordon field on the Politzer spacetime may then proceed as follows.
First, we choose some real $t_0 < \tau_-$ and denote by $\Sigma_{t_0} = \{ (t_0,x) :
x \in \mathbb{R} \}$ the $t = t_0$ hyperplane in $P$ that also appears in Fig.\ 4.
Let $I = I_1 \cup \cdots \cup I_m$ be the union of finitely many intervals at constant coordinate
time, i.e.\ $I_j \subset P$, $I_j = \{ (t_j,x) : a_j < x < b_j \}$ where $t_j \in \mathbb{R}$ and 
$a_j < b_j \in \mathbb{R}$. We define $\mathcal{A}_P(C)$ as the $C^*$-subalgebra of $\mathcal{A}_M(\Sigma_{t_0})$
generated by all ${\bf w}(\phi_P)$ where the $\phi_P$ in $\mathcal{S}_P^0$ emanate from $I$. That means,
$\phi_P$ has Cauchy-data contained in $I$ in a local globally hyperbolic neighbourhood of $\overline{I}$
(this may impose constraints on the relative position of the $I_j$ in order not to be ambiguous). 

Let us discuss this more carefully, but for simplicity 
for the case that $I = \{ (t,x) : a < x < b\}$ is just a single
interval at fixed coordinate-time $t$. We will also need to consider the set $\mathscr{L}_M$ which is
defined to consist of the 8 lightrays (in 2-dim.\ Minkowski spacetime) that pass individually through
one of the boundary points $(\tau,\pm L)$ and $(-\tau,\pm L)$ of the strips $S_+$ and $S_-$.
\\[4pt]
(i) \quad First, suppose that $t \le t_0$. Then let $\phi_P$ be in $\mathcal{S}_P^0$
and let the Cauchy-data of $\phi_P$ restricted to a globally hyperbolic neighbourhood of $\mathcal{I}$ be contained in
$I$. Then we trace the solution $\phi_P$ ``forward in time'' to its Cauchy-data on $\Sigma_{t_0}$. This defines
uniquely a solution $\phi_M$ in $\mathcal{S}_M$, and the ${\bf w}(\phi_P)$ that we have used previously 
is actually ${\bf w}(\phi_M)$, a generating element of $\mathcal{A}_M(\Sigma_{t_0}) = \mathcal{A}(M)$. 
One can see that the thus constructed algebras $\mathcal{A}_P(I)$ coincide with the algebras 
$\mathcal{A}_M(I)$ of the algebras of the quantized real, massless Klein-Gordon field in 2-dim.\ Minkowski
spacetime {\it unless} there is a lightray in $\mathscr{L}_M$  that intersects $I$, 
since the condition $\phi_P \in \mathcal{S}_P^0$ has to be observed.
If those lightrays intersect $I$ e.g.\ in the point $(t,x_\ell)$, then $\mathcal{A}_P(I)$ is the 
$C^*$-algebra generated by $\mathcal{A}_M(I^<)$ and $\mathcal{A}_M(I^>)$ where 
$I^< = \{(t,x) : a < x < x_\ell\}$ and $I^> =\{ (t,x) : x_\ell < x < b\}$.
\\[4pt]
(ii) \quad Now consider the case that $t \ge t_0$. If $I$ is contained in $S_-^\perp$, the causal complement of 
$S_-$ in Minkowski spacetime, then one proceeds as in the case $t \le t_0$, but now tracing 
solutions $\phi_P$ in $\mathcal{S}^0_P$ emanating from $I$ ``backward in time'' to its Cauchy-data
on $\Sigma_{t_0}$ and identifying with the corresponding $\phi_M$ in $\mathcal{S}_M$
This again leads to $\mathcal{A}_P(I) = \mathcal{A}_M(I)$, identified as a $C^*$-subalgebra of
$\mathcal{A}_M(\Sigma_{t_0})$. The procedure is the same for $t < \tau_-$, with the same result.
Again, one must observe a modification if any of the lightrays in $\mathscr{L}_M$  intersects $I$. 
\\[4pt]
(iii) \quad Suppose now that $t > \tau$ and that $I \in J^+(S_+)$, the causal future region of $S_+$
(defined in $M$ and identified as a subregion of $P$). Again, $\phi_P$ with local Cauchy-data supported in
$I$ is ``traced backward in time'' to its Cauchy-data on $\Sigma_{t_0}$, and identified with the 
corresponding solution in $\mathcal{S}_M$. However, due to the second boundary condition of \eqref{condPM},
this is now different from the $\phi_M$ one would obtain by propagating the solution with the 
same data as $\phi_P$ on $I$ backwards in time on Minkowski spacetime: A ``spatial displacement''
is induced through this boundary condition, compared to the propagation on Minkowski spacetime,
if $\phi_P^R$ or $\phi_P^L$ hits on $S_+$. Thus, there is an effect comparable to backward scattering
induced by the boundary condition \eqref{condPM}. 
\\[4pt]
(iv) \quad Finally, we treat the most interesting case of some $I$ with $-\tau < t < \tau$ lying inside
the causal dependence region of $P_{\rm CTC}$. Consider e.g.\ the case of $I$ so that locally, $D(I) = O_2$ 
in Fig.\ 4. Any $\phi_P$ in $\mathcal{S}_P^0$ with local Cauchy-data supported in $O_2$ is a 
superposition of the right-moving part $\phi_P^R$ and left-moving part $\phi_P^R$, as illustrated in Fig.\ 4 (identifying
$\phi_P^L$ with $\tilde{\phi}_P^L$). The Cauchy-data on $\Sigma_{t_0}$ are located in the intervals $C$ and 
$\tilde{C}$ on $\Sigma_{t_0}$. Thus, one can see from Fig.\ 4 that $\mathcal{A}_P(O_2)$ coincides with
the $C^*$-subalgebra of $\mathcal{A}_M(\Sigma_{t_0})$ generated by $\mathcal{A}_M^R(C)$ and 
$\mathcal{A}_M^L(\tilde{C})$. This is different from the algebra one would obtain for
$\mathcal{A}_M(O_2)$: That also coincides with the $C^*$-algebra generated by 
$\mathcal{A}_M(C')$ and $\mathcal{A}_M(\tilde{C}')$ with sub-intervals $C'$ and $\tilde{C}'$ of
$\Sigma_{t_0}$, but they are different (spatially displaced) from $C$ and $\tilde{C}$. Furthermore,
Fig.\ 4 also shows that
\begin{equation} \label{algebra-regions}
 \begin{split}
  \mathcal{A}_P(O_2) & = \mathcal{A}_P(O_j) \quad (j = 1,3,4) \\
                     & = \mathcal{A}_P(\text{any of the purple, triangular regions}) \\
                     & = C^*\text{-algebra generated by the} \ \mathcal{A}_P(O_k)\,,\ k=1,2,3,4 \\
                     & = \mathcal{A}_P \left( \bigcup_{k = 1,2,3,4} O_k \cup (\text{all the purple, triangular regions})\right)
 \end{split}
\end{equation}
Thus, also in this case, $\phi_P$ with local Cauchy-data in $O_2$ has different ``traced backwards in time'' Cauchy-data
on $\Sigma_{t_0}$ in the presence of the first constraint in \eqref{condPM} than without that constraint, so
that again one has some backward scattering effect induced by the constraints \eqref{condPM}. In view of 
Fig.\ 4, our use of ``local Cauchy-data'' has been somewhat ambiguous --- one would have to more precisely
define the ``size'' of an ambient globally hyperbolic neighbourhood of $\overline{I}$ --- but then \eqref{algebra-regions}
shows that this does not lead to inconsistencies when it comes to the definition of local algebras. The 
point to emphasize, however, is that observables in $\mathcal{A}_P(O_2)$ are {\it not localized in $O_2$ in
the same sense as in Minkowski spacetime}. In fact, the observables in $\mathcal{A}_P(O_2)$ are not more sharply
localized than in 
\begin{equation}
 \tilde{O} = \bigcup_{k = 1,2,3,4} O_k \cup (\text{all the purple, triangular regions})
\end{equation}
Moreover, $\mathcal{A}_P(O_2)$ and, say, $\mathcal{A}_P(O_3)$ really are {\it identical}. That means the 
algebra generated by $\mathcal{A}_P(O_2)$ and $\mathcal{A}_P(O_3)$ is not in any sense a copy 
of $\mathcal{A}_P(O_2) \otimes \mathcal{A}_P(O_3)$ where one has elements ${\bf a}_2 \otimes {\bf a}_3$
corresponding to carrying out observable ${\bf a}_2$ in $O_2$ and observable ${\bf a}_3$ in $O_3$ independently.
The algebra generated by $\mathcal{A}_P(O_2)$ and $\mathcal{A}_P(O_3)$ enforces carrying out the same
observable ${\bf a}$ in $O_3$ upon carrying out ${\bf a}$ in $O_2$, and vice versa (very roughly, one always has 
${\bf a}_2 = {\bf a}_3$). This circumstance leads to severe constraints on trying to use 
quantum field theory on the Politzer spacetime in attempts to circumvent the quantum no-cloning theorem,
and we hope to return to a discussion of that issue elsewhere. 

On enlarging $O_2$, one again has to modify the definition if the lightrays emanating from the boundary points
$(\tau,\pm L)$ or $(-\tau,\pm L)$ intersect $O_2$ (these lightrays re-appear periodically inside $P_{\rm CTC}$)
in a similar manner as mentioned in (i) above. 
\\[10pt]
The construction of the $\mathcal{A}_P(O)$ leads to a family of $C^*$-subalgebras of $\mathcal{A}_M(\Sigma_{t_0}) =
\mathcal{A}(M)$ satisfying isotony. The locality condition would have to be modified; also, one should take
into account that some algebras $\mathcal{A}_P(O)$ actually agree with algebras corresponding to regions which are larger
than $O$, and possibly disconnected. This is a feature which does not occur in the same manner for 
arbitrarily localizable quantum field theory in globally hyperbolic spacetimes. An interesting feature is the 
possibility to describe the effect of the constraints \eqref{condPM} in terms of a scattering transformation
as also discussed in a similar context for classical fields e.g.\ in \cite{FriedmanMorris}. Causality-violating
wave propagation on non-commutative spaces can also be described in terms of scattering transformations
\cite{LechnerVerch} and it would be of interest to investigate relations between these approaches and the question
if the local algebras $\mathcal{A}_P(O)$ can be reconstructed from sufficient information about the 
scattering transformation induced by the boundary constraints \eqref{condPM}. 

It is not difficult to see that our construction of the local algebras $\mathcal{A}_P(O)$ of the quantized
real, massless Klein-Gordon field on the Politzer spacetime is independent of the choice of $t_0$ for 
$\Sigma_{t_0}$ as long as $t_0 < -\tau$; effectively, this is a consequence of the time-translation covariance
of the theory on Minkowski spacetime. In our construction, the local algebras $\mathcal{A}_P(O)$ agree
with the $\mathcal{A}_M(O)$ for $O$ in $S_-{}^\perp \cup \{(t,x) : t < -\tau,\ x \in \mathbb{R}\}$, and differ
from the $\mathcal{A}_M(O)$ if $O$ lies outside this region, i.e.\ in the causal influence domain of $P_{\rm CTC}$.
Hence, our construction resembles the idea of a ``time machine set into operation at some earliest time $-\tau$''
as discussed in \cite{Hawking,EarmanSmeenkWuethrich}. The boundary of $S_-{}^\perp \cup \{(t,x) : t < -\tau,\ x \in \mathbb{R}\}$
would then correspond to the {\it Cauchy horizon} of $\Sigma_{t_0}$ to the past of which the Politzer spacetime 
is globally hyperbolic. Not having completely settled the differentiable structure of Politzer spacetime, it is 
difficult to precisely define the Cauchy horizon of $\Sigma_{t_0}$, but a reasonable (minimal) choice is depicted in
Fig.\ 5:
\begin{center}
 \includegraphics[width=12cm]{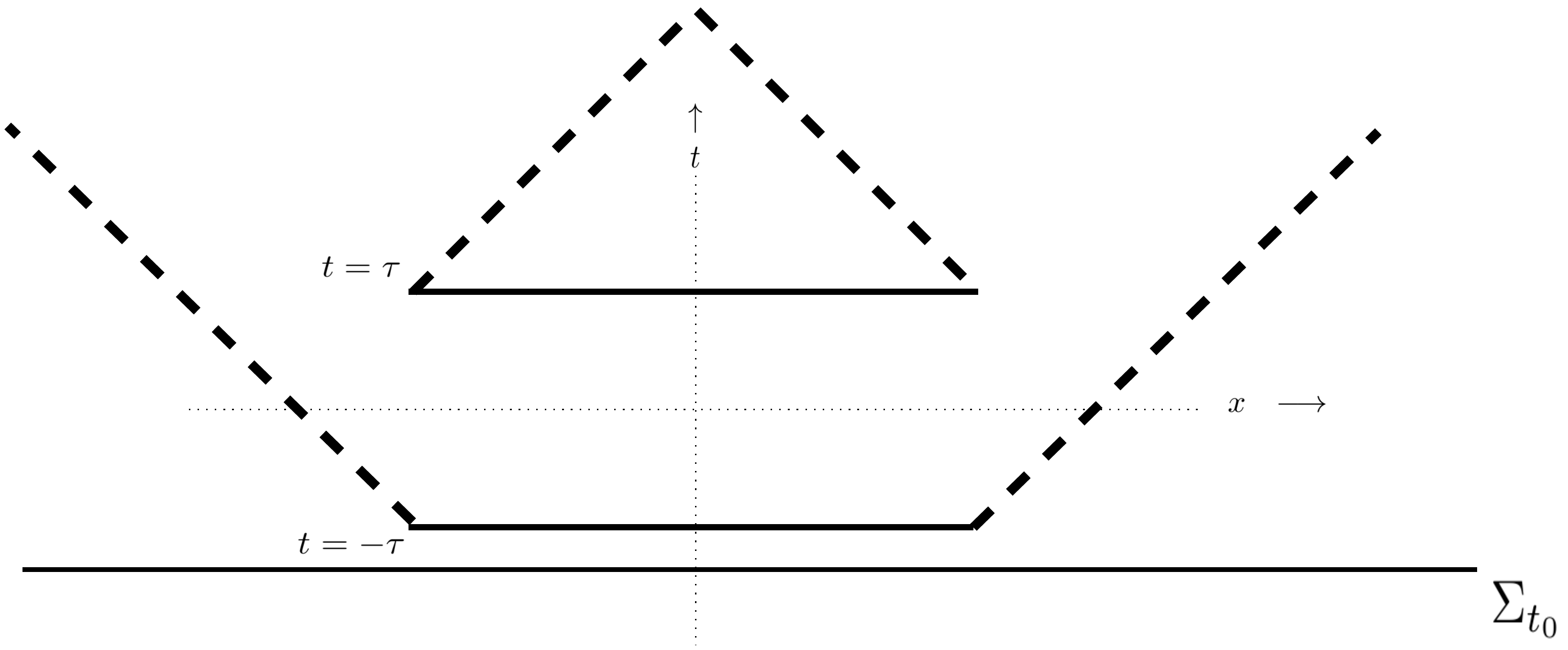}
\end{center}
{\small
{\bf Figure 5}. The thick broken lines indicate the Cauchy horizon of $\Sigma_{t_0}$. 
}
\\[8pt]
This particular form of the Cauchy horizon is not compactly generated --- its past inextendable generators are 
the indicated lightrays that ``end'' at the boundary points of the $S_\pm$ which do not belong to the spacetime,
so the past inextendable generators are not past-confined to any compact set within the Politzer spacetime.
Nevertheless, the idea of \cite{Hawking} that the ``time machine region'' is of finite extension 
is probably still well met, and likewise the boundary condition \eqref{cond-0} ensures that there is 
nothing flowing into the system from the boundary points of the $S_\pm$, which was the motivation for
considering compactly generated Cauchy horizons in \cite{Hawking}. The failure of the Cauchy horizon to 
be compactly generated renders the arguments of \cite{KayRadWald} against quantum field theory on spacetimes
with a compactly generated Cauchy horizon inapplicable. However, one could potentially alter the 
spacetime structure such that another version of the Politzer spacetime, and of a quantum field theory
constructed on it, allow the arguments of \cite{KayRadWald} to be applied. Essentially that would then state
that such a Politzer spacetime, and quantum field theory on it, is likely to be unstable as a solution to
the semiclassical Einstein equation, or even ``too singular'' to be considered as a solution to the semiclassical
Einstein equation. Yet, this is a different level of discussion of quantum fields on spacetimes with
CTCs which is beyond the scope of this work. 
\\[6pt]
An objection against our construction of the local observable algebras $\mathcal{A}_P(O)$ on the Politzer
spacetime could be that we have a built-in time-asymmetry since, as mentioned, the local observables 
agree ``in the past of the Cauchy horizon'' with the local observables $\mathcal{A}_M(O)$ on Minkowski
spacetime but differ to the future of the Cauchy horizon. This is due to having constructed 
the $\mathcal{A}_P(O)$ with reference to some hyperplane $\Sigma_{t_0}$ where $t_0 < -\tau$. Thereby an
element is introduced into the construction of a quantum field theory on Politzer spacetime in
addition to $P$ and the boundary conditions \eqref{condPM} and \eqref{cond-0}, in such a way that the 
time-orientation symmetry respected by $P$, \eqref{condPM} and \eqref{cond-0}, is broken. 
A way to avoid this is e.g.\ to ``double'' the system and consider the algebras $\mathcal{A}_P(O)$ constructed
with respect to $\Sigma_{t_0}$ and $\mathcal{A}_{\overline{P}}(O)$ constructed with respect to $\Sigma_{-t_0}$ 
simultaneously. Another way is to construct the $\mathcal{A}_P(O)$ similarly as before, but with
respect to $\Sigma_{t_0 = 0}$, the $t = 0$ hyperplane. One can see here that for a spacetime with CTCs,
there may be several different quantum field theories that can be constructed and be regarded 
as ``quantization'' of the same classical field model.
\\[6pt]
As another feature, our construction of the $\mathcal{A}_P(O)$ fulfills the 
{\it F-locality} condition of Kay \cite{KayFlocality}. The F-locality condition demands, in the present
context, that for any point $p = (t,x)$ in the Politzer spacetime $P$ there should be an open, globally
hyperbolic neighbourhood $O_p$ of $p$ (in $P$) such that there is an isomorphism
$\chi : \mathcal{A}_M(O_p) \to \mathcal{A}_P(O_p)$ so that $\chi(\mathcal{A}_M(O)) = \mathcal{A}_P(O)$
for all $\overline{O} \subset O_p$. The exception is that $p$ must not lie on any lightray emanating
from the boundary points $(\tau,\pm L)$ or $(-\tau,\pm L)$ of $S_\pm$. On the other hand, there is
a vacuum state for the free, massless Klein-Gordon field with corresponding vacuum GNS Hilbert
space representation of $\mathcal{A}(M)$.\footnote{At this point, the remark of the previous
footnote becomes relevant so as to
avoid infrared problems in the vacuum representation.} 
We anticipate that
the local von Neumann algebras $\mathcal{N}_P(O)$ in the vacuum Hilbert space representation
have the property  $\mathcal{N}_P(D(I_p)) = \mathcal{N}_P(D(I))$ where
$I$ is an open equal-time interval containing $p$, and $I_p = I \backslash \{p \}$. Given
that this property holds, F-locality would hold at the level of local von Neumann algebras
in the vacuum representation. 
\\[6pt]
In the articles \cite{KayFlocality,FewsHiguchi}, systems of local algebras of observables 
have been constructed for the quantized real, massless Klein-Gordon field on the two-dimensional
spacetime $S^1 \times \mathbb{R}$ with a ``rolled-up'' time-axis which contains CTCs. The construction is based 
on defining counterparts of the ``advanced-minus-retarded Green's function'' on this spacetime;
on any globally hyperbolic spacetime $M$, such a Green's function is uniquely determined by the field
equation and defines the symplectic form $\sigma_M$ and thereby the Weyl-algebra on which 
the construction of the family of local algebras is based. One can view $P_{\rm CTC}$ as 
a part of the $S^1 \times \mathbb{R}$ spacetime and thus it is likely that our construction of the 
quantized real, massless Klein-Gordon field corresponds to a choice of ``advanced-minus-retarded Green's function''
on $S^1 \times \mathbb{R}$ and, after restriction to $P_{\rm CTC}$, choice of an extension to
the full Politzer spacetime. 
We view an analysis of such a connection as a promising line of further investigation.

\section{Conclusion}   \setcounter{equation}{0}

We have obtained complementary results on the possibility of realizing the D-CTC condition in quantum field theory.
None of our results is a clear ``no-go'' or ``always go'' statement, but leaves some room for interpretation, depending
on whether one is inclined to take the D-CTC condition e.g.\ in the form \eqref{DC} as characterizing for causality violating quantum
processes --- i.e., involving CTCs --- or not. Let us begin with our first result, Prop.\ \ref{prop1}.
It is totally reasonable to require that the solution state $\tilde{\omega}$ and the state
$\tilde{\omega}(U^\ast\,.\,U)$ should have finite total energy, and consequently, they would fulfill
the analyticity assumptions of Prop.\ \ref{prop1}. From this vantage point, the ensuing negative result can 
then be interpreted as stating
that within the framework of localized subsystems in (stationary) globally hyperbolic spacetimes, the
possiblity of finding solutions to the D-CTC condition is severely constrained.
Similarly, if one takes the point of view that a prime distinguishing feature of quantum physics as opposed to
classical physics consists in entangled states between acausally related system parts and therefore expects that
solution states $\tilde{\omega}$ to the D-CTC condition \eqref{DC} should satisfy the Reeh-Schlieder
property (so that they show strong acausal entanglement),
then Prop.\ \ref{prop2} can be viewed as an obstacle for that expectation to ever become realized, as long 
as the correlations in $\tilde{\omega}$ and $\tilde{\omega}(U^*\,.\,U)$ stay comparable. 
Proponents of the point of view that the
D-CTC condition should be seen as characterizing for the presence of processes that are based on the occurrence of CTCs might take 
these results as (indirectly) 
supporting their position, but we should point out that this is not compelling, in particular as 
no specifications of the behaviour of $U$ with respect to the locality structure of the given quantum
field theory are involved.

In contrast, the result of Prop.\ \ref{approxiDeutschinQFT} states that whatever $U$ and the partial state $\omega_A$,
approximate solution states $\tilde{\omega}$ to the D-CTC condition \eqref{DC} can always be obtained in quantum field theory 
on {\it globally hyperbolic spacetimes} (without any involvement of CTCs in the sense of general relativity) under very
general assumptions. Sceptics towards the position that the D-CTC condition is characteristic for the occurence of 
CTC-based processes
will obviously take this as a strong argument in favour of their case. Yet, there remains the possibility that in quantum field 
theory on globally hyperbolic spacetimes, the D-CTC condition cannot be fulfilled exactly under the assumptions of 
Prop.\ \ref{approxiDeutschinQFT}, while there could be models of quantum fields on spacetimes containing CTCs which 
do admit exact solutions to the D-CTC problem under ``similar'' assumptions. That is an interesting question, deserving
further investigation.

However, we conclude that, in particular in view of Prop.\ \ref{approxiDeutschinQFT}, it is very difficult to judge if the 
D-CTC condition \eqref{DC} can really say very much about quantum processes in the presence of CTCs in the sense
of general relativity. To this end, it is not sufficiently specific, at least within the framework of quantum field
theory in globally hyperbolic spacetimes. This refers in particular to the unitary operator $U$: Information how its action relates to
the locality and causality structure of the quantum field theory under investigation would have to be supplied.
Therefore, we think that statements like ``quantum mechanics therefore allows for causality violation without paradoxes 
whilst remaining consistent with
relativity'' \cite{RingbauerEtAl} --- apparently based on viewing the D-CTC condition as characteristic for the 
presence of processes involving CTCs --- should be taken with great caution. The D-CTC condition originates from
quantum communication networks (quantum circuits) without any relativity related context. In contrast,
such a context is
provided by the approach to quantum field theory due to Haag and Kastler \cite{HaagKastler,Haag} 
which uses concepts of locality, causality and covariance as fundamental ingredients. Our construction
of the quantized real, massless Klein-Gordon field on the Politzer spacetime shows that the Haag-Kastler
framework is general enough to describe quantum physics on certain spacetimes with CTCs, even though the  concept 
of localization acquires a character which is different from the concept of arbitrary localizability of
quantized fields on globally hyperbolic spacetimes. A natural task for future research will be 
to investigate the D-CTC condition within the framework of quantum field theory on the Politzer spacetime, and
on other types of spacetimes containing CTCs.
\\[24pt]
{\bf Acknowledgements} J.T.\ would like to thank the MPI for Mathematics in the Sciences for financial support.
Both authors would like to thank Felix Finster for discussions on the topic, particularly related to 
hyperbolic PDEs on spacetimes with CTCs. R.V.\ would like to thank Chris Fewster for helpful conversations
on QFT in spacetimes with CTCs related to content of this work, 
as well as Detlev Buchholz and Christian Wuethrich for response to questions.

\section{Appendix}    \setcounter{equation}{0}

In this Appendix, we show (for the sake of completeness, since the result is well-known,
but not easy to trace in the literature in a form relating to a simple quantum 
system) that there is a state on ${\sf B}(\mathcal{H})$ 
(regarded as $C^\ast$ algebra) for an infinite-dimensional, separable
Hilbert space $\mathcal{H}$ which is not given by a density matrix. Furthermore, this state
arises as weak-$\ast$ limit of density matrix states. 

We can take $\mathcal{H} =L^2(\mathbb{R})$ and view it as the Hilbert space of the harmonic oscillator
in one space dimension. For the harmonic oscillator, there is a selfadjoint Hamilton operator 
${\bf h}$ which has an eigenvalue spectrum ${\rm spec}({\bf h}) = 
\{ E_n = n + \frac{1}{2} : n \in \mathbb{N}_0\}$.
There is an up to a phase unique unit-norm eigenvector $\psi_n$ for each eigenvalue $E_n$. Given
any $\beta>0$, there is the Gibbs-state at inverse temperature $\beta$, given by
\begin{equation}
 \omega_\beta({\bf a}) = \frac{1}{Z_\beta}{\rm Tr}({\rm e}^{-\beta {\bf h}}{\bf a}) \quad \ \ \ ({\bf a} \in
 {\sf B}(\mathcal{H}))\,.
\end{equation}
Here, $Z_\beta = {\rm Tr}({\rm e}^{-\beta {\bf h}}) = \sum_n {\rm e}^{-\beta E_n}$ is the partition function.
The Banach-Alaoglu Theorem \cite{ReedSimon1} asserts the weak-$\ast$ compactness of the state space of
${\sf B}(\mathcal{H})$ viewed as a $C^\ast$ algebra. This applies in particular to the 
family of states $\omega_\beta$, $0 < \beta < 1$. Consequently, there is a net (generalized sequence)
$\{\beta_\kappa\}_{\kappa \in K}$, where $K$ is some directed set, with $\lim_\kappa \beta_\kappa = 0$,
so that $\omega_{\beta_\kappa}({\bf a})$ converges for all ${\bf a} \in {\sf B}(\mathcal{H})$,
and such that
\begin{equation}
 \omega'({\bf a}) = \lim_\kappa \,\omega_{\beta_\kappa}({\bf a}) \quad \ \ \ ({\bf a} \in {\sf B}(\mathcal{H}))
\end{equation}
defines a state $\omega'$ on the $C^\ast$ algebra ${\sf B}(\mathcal{H})$. 
Clearly, it holds that 
\begin{equation}
 \omega'({\bf 1}) = \lim_\kappa\, \omega_{\beta_\kappa}({\bf 1}) = 1\,.
\end{equation}
On the other hand, if ${\bf p}$ is any finite-dimensional projector in ${\sf B}(\mathcal{H})$, then
\begin{equation} \label{projecvanish}
 \omega'({\bf p}) = 0\,.
\end{equation}
To see this, note that 
\begin{equation}
 \omega_\beta({\bf p}) = \frac{1}{Z_\beta} \sum_{n = 0}^\infty {\rm e}^{-\beta E_n}
 \langle \psi_n,{\bf p} \psi_n \rangle\,.
\end{equation}
Since ${\bf p}$ is a finite-dimensional projector, it holds that
\begin{equation}
 s = \sum_{n = 0}^\infty \langle \psi_n, {\bf p} \psi_n \rangle  < \infty\,.
\end{equation}
As $\lim_\kappa\, \beta_\kappa = 0$, it follows that $\lim_\kappa\, Z_{\beta_\kappa} = 
\lim_{\kappa}\,{\rm e}^{-\beta_\kappa /2}/(1 - {\rm e}^{-
\beta_\kappa}) = \infty$.
Therefore, for any given $\varepsilon > 0$ there is some $\kappa \in K$ such that
\begin{equation}
  \frac{s}{Z_{\beta_{\kappa'}}} < \varepsilon \quad \text{for all}\ \ \kappa' \succ \kappa
\end{equation}
where $\succ$ is the ordering relation on the index set $K$. This implies that
\begin{equation}
 \omega_{\beta_{\kappa'}}({\bf p}) = \frac{1}{Z_{\beta_{\kappa'}}}
 \sum_{n=0}^\infty {\rm e}^{-\beta_{\kappa'}E_n}
  \langle \psi_n, {\bf p} \psi_n\rangle \le \frac{s}{Z_{\beta_{\kappa'}}} < \varepsilon
\end{equation}
for all $\kappa' \succ \kappa$. This proves \eqref{projecvanish} for all finite-dimensional projectors
${\bf p}$. 
On the other hand, assume --- by contradiction --- that there was a density matrix 
$\varrho'$ on $\mathcal{H}$ so that $\omega'({\bf a}) = {\rm Tr}(\varrho' {\bf a})$ holds
for all ${\bf a} \in {\sf B}(\mathcal{H})$. Making use of the spectral decomposition of
$\varrho'$, there is a sequence of non-negative real numbers $r_n$, $n \in \mathbb{N}_0$,
with $\sum_n r_n = 1$, and an orthonormal basis $\{ \eta_n\}_{n \in \mathbb{N}_0}$ of
$\mathcal{H}$, so that 
\begin{equation}
 {\rm Tr}(\varrho' {\bf a}) = \sum_{n=0}^\infty r_n \langle \eta_n,{\bf a} \eta_n \rangle\,.
\end{equation}
Setting ${\bf p}_k = \sum_{n=0}^k |\eta_n\rangle \langle \eta_n|$, one obtains
\begin{equation}
\lim_{k \to \infty}\, {\rm Tr}(\varrho' {\bf p}_k) = 1\,,
\end{equation}
but having assumed $\omega'({\bf a}) = {\rm Tr}(\varrho' {\bf a})$ for any bounded operator
${\bf a}$, this contradicts \eqref{projecvanish}. Thus, $\omega'$ is a state on 
${\sf B}(\mathcal{H})$ which is not given by a density matrix.
${}$ \hfill $\Box$

\end{document}